\documentclass[a4paper,fleqn,usenatbib]{mnras}

\usepackage{newtxtext,newtxmath}

\usepackage[T1]{fontenc}
\usepackage{ae,aecompl}

\usepackage{graphicx}	
\usepackage{subfig}
\usepackage{amsmath}	
\usepackage{amssymb}	

\title[A~proper motion study]{A~ground--based proper motion study of twelve nearby 
Globular Clusters}

\author[W. Narloch et al.]{
W. Narloch$^{1}$\thanks{E-mail: wnarloch@camk.edu.pl (WN); mnr@camk.edu.pl (MR); poleski.1@osu.edu (RP); 
pych@camk.edu.pl (WP); ian@obs.carnegiescience.edu (IBT)},
J. Kaluzny\thanks{Deceased},
R. Poleski$^{2}$,
M. Rozyczka$^{1}$,
W. Pych$^{1}$, 
I. B. Thompson$^{3}$
\\
$^{1}$Nicolaus Copernicus Astronomical Center, of the Polish Academy of Sciences, ul. Bartycka 18, 00-716 Warsaw, Poland\\
$^{2}$Department of Astronomy, Ohio State University, 140 West 18th Avenue, Columbus, OH 43210, USA\\
$^{3}$The Observatories of the Carnegie Institution of Washington, 813 Santa Barbara Street, Pasadena, CA 91101, USA
}

\date{Accepted XXX. Received YYY; in original form ZZZ}

\pubyear{2016}

\begin{document}
\label{firstpage}
\pagerange{\pageref{firstpage}--\pageref{lastpage}}
\maketitle

\begin{abstract}
We derive relative proper motions of stars in the fields of the globular clusters M12, NGC~6362, 
M4, M55, M22, NGC~6752, NGC~3201, M30, M10, NGC~362, M5, and 47~Tucanae based on data collected 
between $1997$ and $2015$ with the $1$-m Swope telescope of Las Campanas Observatory. 
We determine membership class and membership probability for over $446\,000$ objects, and show 
that these are  efficient methods for separating field stars from members of the cluster. 
In particular, membership probabilities of variable stars and blue/yellow/red stragglers 
are determined. Finally, we find absolute proper motions for six globular clusters from our sample: 
M55, NGC~3201, M10, NGC~362, M5, and 47~Tuc. An electronic catalogue of the derived proper motions 
is publicly available via the internet.
\end{abstract}




\section{Introduction}

Globular clusters (GCs) are important laboratories to study both stellar evolution and dynamics 
as well as chemical evolution of the universe. The analysis presented in this work is a~part of 
the CASE project \citep[Cluster AgeS Experiment,][]{kaluzny2005case} which is devoted to a~search 
for and follow-up observations of variable stars in the fields of nearby GCs, in particular, for 
detached eclipsing binaries which might be useful in determining distances and ages of GCs 
\citep[e.g.][]{thompson2010}. An additional result of the project is the discovery of a~large number 
of previously unknown variable stars of other types found in the fields of the GCs, for example 
RR~Lyrae stars which might be interesting for astroseismic studies \citep{smolec2017}. But to fully 
benefit such studies of the properties of GCs, it is important to separate members of the clusters 
from field stars. Reliable separation can be achieved, e.g., by a~proper motion (PM) study. 

The first PM measurements in GCs were made using photographic plates 
\citep[][and references therein]{ebbighausen1942,cudworth1980}. Later studies replaced the photographic 
plates with CCD detectors. The CCD data allowed a significant reduction in the level of statistical 
uncertainties and  permitted reliable results to be obtained for a larger number of fainter stars 
even with a~short time base of only a~few years \citep{anderson2006,bellini2009}. 

The launch of the Hubble Space Telescope (HST) enabled studies of dense central parts of GCs, and 
PM measurements of stars with a~high accuracy (tens of $\mu$as/yr for the brightest stars). Such 
precision  is crucial in, e.g., studies of internal kinematics of GCs \citep[e.g.][]{watkins2015}. 

The GAIA space mission is expected to provide astrometry for about $1$~billion point sources 
down to $V\approx$20~mag with precision of tens of $\mu$as or better for objects brighter than 
$V\approx$15~mag, and radial velocities down to $V\approx$17~mag \citep{mignard2005,pancino2013,
pancino2017}. It aims to chart a~three--dimensional map of the Milky Way out to a~distance of 
$\approx$8.5~kpc from the Sun including fields of all GCs considered in this work. The first 
data release occurred last year \citep{lindegren2016}, and contained high-accuracy positions for 
more than $1$~billion stars brighter than $V\approx21$~mag. This limit is still about one magnitude 
shallower than what we have obtained in this work. We have thus been able to measure more stars 
than GAIA in the fields of our sample of GCs (especially close to the cluster center).

At this moment, GAIA PMs have been provided for only the brightest stars. This is in common with 
the Hipparcos and Tycho-2 catalogues, which omit the fields of GCs. This is a~distinct advantage 
of our catalogue comparing to the first release of the GAIA catalogue. The expected final 
accuracy of GAIA PMs for GC members, $\approx$1~mas/yr at $V$~$\approx$21~mag, is comparable to ours 
(see Section \ref{sec:error}). For brighter stars GAIA measurements will be more accurate (e.g. 
$\approx$20 vs. $\approx$60 $\mu$as/yr at $V\approx$15~mag); however the final accuracy is unlikely 
to be reached before 2020, and only the latest few Gaia releases (2020--2023) are expected to provide  
a~significant breakthrough in GC research \citep{pancino2017}.

In this paper we present studies of relative proper motions of stars in the fields of twelve nearby 
galactic GCs: M12, NGC~6362, M4, M55, M22, NGC~6752, NGC~3201, M30, M10, NGC~362, M5, and two fields 
(East and West) of 47~Tuc. Previous analyses of observations of the first six GCs was published by 
\citet{zloczewski2011,zloczewski2012} and for M55 by \citet{sariya2012}. In the first two papers the 
authors used data from $2.5$~m du Pont telescope with a~field of view of $8.83 \times 8.83$~arcmin$^2$. 
Although they obtained reliable PMs and membership status for stars close to the center of the cluster, 
these studies lack information about PMs of more distant objects. Data for the third paper were 
collected with the Wide Field Camera (WFI) mounted on the $2.2$~m MPG/ESO telescope. The authors 
created a~membership probability catalogue in a~wide field ($26 \times 22$~arcmin$^2$), however the 
faintest stars they measured were no fainter than $\approx$20~mag in $V$-band, comapared to 
$\approx$21.5~mag in the present work. In addition, the camera had some gaps between CCD chips,and so 
objects located in these regions were not measured. A~PM catalogue for M10 based on photographic plates 
and Hipparcos data was presented by \citet{chen2000}, where they provided absolute PMs for $532$ stars. 
The recent study of \citet{cioni2016} resulted in a~catalogue of absolute PMs in 47~Tuc based on VISTA 
data. With a~time baseline of just $1$--yr they measured PMs of about $86\,000$ stars located 
$10-60$~arcmin from the cluster center, preferentially in the direction to the Small Magellanic Cloud 
(SMC, i.e. in a~field partially covering that of ours). Extensive studies of PMs in GCs from our list 
have also been made with the HST, 
e.g., \citet[][for NGC~6752]{drukier2003}, \citet[][for 47~Tuc]{mclaughlin2006}, 
\citet[][in regions of $22$ galactic GCs]{bellini2014} or \citet[][for $38$ galactic GCs]{simunovic2016}. 
HST data provide deep and precise photometry for thousands of stars impossible to resolve with 
ground--based telescopes, but the data cover a~field of view of only about $3 \times 3$~arcmin$^2$, 
which is much smaller than in the present analysis. Thus, in many aspects, the work presented here 
is a~valuable supplement to the surveys quoted above. 

The absolute PMs of GCs are important quantities, which when combined with radial velocities of the 
systems allow a derivation of the space motions of GCs. Knowledge of the latter helps to build dynamical 
models of the Galaxy, trace the gravitational potential of the Galaxy, or find the origin of the GCs 
themselves. Presently, for each GC considered in this work, there are at least a~few measurements of 
absolute PMs. The earliest results for all our GCs but one are summarized in \citet{dinescu1999}, later 
calculations come from e.g. \citet{chen2000,dambis2006,casetti-dinescu2007} or \citet{zloczewski2011}. 
Only 47~Tuc has several measurements \citep[except already cited, also][]{freire2001,freire2003,anderson2003}. 
The newest ones come from \citet{cioni2016} and \citet{watkins2016}. However, all these determinations 
are based on different methods, and produce different results. For this reason it is valuable to make 
additional independent measurements, which is what we have done in the present work.

The paper is organized as follows. In Section~\ref{sec:data} we describe observational data selection 
and reduction. Procedures employed for measuring PMs and calculating membership probabilities of 
individual stars are discussed in Section~\ref{sec:pm}. Section~\ref{sec:cmd} contains color--magnitude 
diagrams (CMDs) for the analyzed cluster fields. In Section~\ref{sec:abspm} we derive absolute PMs 
for six GCs from our sample, and Section~\ref{sec:summary} provides a~brief summary of the paper. 

\section{Data selection and preparation}
\label{sec:data}

The images analyzed in this paper were collected within the CASE project between the years $1997-2015$. 
Observations in $V$ and $B$ filters were obtained using the $1$-m Swope telescope located at  Las 
Campanas Observatory in Chile. Two CCD cameras were used: SITe\#3 with a~field of view of $14.8 
\times 22.8$ arcmin$^2$ and a~scale of $0.435$ arcsec/pixel during years $1997-2010$ and E2V~CCD231-84 
with a~field of view of $29.7 \times 29.8$ arcmin$^2$ and the same pixel scale in $2015$. Equatorial 
coordinates of the field centers of the reference images are listed in Tab.~\ref{tab:eqcoo}.

The CASE project was not designed for astrometry purposes and in order to save  CCD readout time 
a~subraster was often used. As a result the final field sizes of our data sets were not uniform. 
As a~consequence, the PM errors are larger for stars located close to the edge of a~reference image, 
because of the smaller number of epochs used in the calculations. Tab.~\ref{tab:frames} contains 
information about our data sets.

\begin{table*}
	\centering
	\caption{Equatorial and galactical coordinates of the centers of reference images.}
	\label{tab:eqcoo}
	\begin{tabular}{llrrrr} 
		\hline
		Field & Messier & $\alpha_{2000}[^\circ]$ & $\delta_{2000}[^\circ]$ & $l[^\circ]$ & $b[^\circ]$ \\
		\hline
		NGC~6218  & M12      & $251.824072$ &  $-1.984139$ &  $15.69008$ &  $26.28193$\\
		NGC~6362  &          & $262.982455$ & $-67.032649$ & $325.56947$ & $-17.56350$ \\
		NGC~6121  & M4       & $245.911419$ & $-26.524932$ & $350.98267$ &  $15.96286$ \\
		NGC~6809  & M55      & $294.999089$ & $-30.963256$ &   $8.79412$ & $-23.27151$ \\
		NGC~6656  & M22      & $279.103212$ & $-23.913256$ &   $9.88592$ &  $-7.55835$ \\
		NGC~6752  &          & $287.712680$ & $-59.985725$ & $336.49111$ & $-25.62647$ \\
		NGC~3201  &          & $154.400593$ & $-46.411364$ & $277.22645$ &   $8.64011$ \\
		NGC~7099  & M30      & $325.117975$ & $-23.180452$ &  $27.18754$ & $-46.85866$ \\
		NGC~6254  & M10      & $254.290704$ &  $-4.100178$ &  $15.13883$ &  $23.07343$ \\
		NGC~362   &          &  $15.818869$ & $-70.849659$ & $301.52851$ & $-46.24641$ \\
		NGC~5904  & M5       & $229.649514$ &   $2.079606$ &   $3.86727$ &  $46.78676$ \\
		NGC~104~E & 47~Tuc~E &   $6.372329$ & $-72.086153$ & $305.74390$ & $-44.90031$ \\
		NGC~104~W & 47~Tuc~W &   $5.661052$ & $-72.085975$ & $306.04922$ & $-44.86778$ \\
		\hline
	\end{tabular}
\end{table*}

\subsection{Data reduction}
\label{sec:data_reduction}

For each GC reference images in both $V$ and $B$ filters were prepared. These were used for CMD 
construction and compilation of a~list of reference stars (hereafter: master list). The reference 
images were selected to have the best possible quality and to cover the widest possible field of 
view. To that end, we chose a~few to several best images obtained consecutively during one night 
with the same exposure time, low air masses, and background, and stacked them into one averaged 
image using the Difference Image Analysis PL (DIAPL) package\footnote{Originally written by 
\citet{wozniak2000} and developed by W.~Pych. Available at http://users.camk.edu.pl/pych/DIAPL/}. 
The number of stacked images varied from $7$ to $18$ in the $V$ filter and from $4$ to $8$ in the 
$B$ filter (in the case of M5 only a~single exposure in the $B$ filter was used). The stacked images 
were cleaned of cosmic rays and bad pixels, and have a~significantly higher signal to noise ratio 
than any single exposure. To reduce the effects of PSF variability, each reference image 
was divided into overlapping subframes (usually $24$ but in the case of M30 and NGC~362, which are 
strongly concentrated, only $6$), which were analyzed independently. Profile photometry 
for each subframe was measured with the DAOPHOT/ALLSTAR package \citep{stetson1987} assuming 
a~Gaussian function with spatial variability to characterize the PSF. Because of crowding, master 
lists were obtained iteratively, gradually decreasing the detection threshold. In the final iteration 
the images were examined by eye and stars omitted in the automatic procedure were added manually. 
In the end, aperture corrections, obtained for each subframe separately using the DAOGROW 
package \citep{stetson1990}, were applied.
Instrumental CMDs were then derived from these photometry files.

PMs were calculated based on individual images in the $V$-filter selected from among all 
available exposures of a~given cluster (see Table~\ref{tab:frames}). For further analysis frames 
characterized by seeing ranging from $\approx$1.17 to $\approx$1.52~arcsec (slightly larger only for 
47~Tuc), the lowest possible background, and air mass lower than $1.45$ were used. The latter 
constraint is a~compromise between minimizing refraction effects and choosing the maximal number of 
single frames used for the calculations. These were then divided into the same number of subframes 
as the corresponding reference images.

Subsequently, stars from the reference lists were identified in each subframe of a~given cluster, 
and profile photometry was measured  with the ALLSTAR parameter REDET set to $1$, enabling 
a~re--determination of star coordinates.

\begin{table*}
	\centering
	\caption{Information about data sets used for PM calculations.}
	\label{tab:frames}
	\begin{tabular}{lccccc} 
		\hline
		Field & number & number & time span between & mean & standard deviation\\
		 & of epochs & of frames & first and last epoch [yr] & FWHM [arcsec] & of mean FWHM [arcsec]\\
		\hline
		M12      &  $8$ &  $779$ &  $9.12$ & $1.31$ & $0.11$ \\
		NGC~6362 & $11$ & $1233$ & $10.40$ & $1.39$ & $0.12$ \\
		M4       & $11$ & $1640$ & $11.06$ & $1.32$ & $0.14$ \\
		M55      & $12$ & $2141$ & $12.37$ & $1.40$ & $0.18$ \\
		M22      &  $5$ & $1254$ &  $8.35$ & $1.32$ & $0.11$ \\
		NGC~6752 &  $3$ &  $405$ & $12.32$ & $1.38$ & $0.10$ \\
		NGC~3201 &  $6$ &  $579$ & $11.21$ & $1.32$ & $0.11$ \\
		M30      &  $2$ &  $531$ & $14.98$ & $1.31$ & $0.10$ \\
		M10      &  $3$ &  $764$ & $17.06$ & $1.31$ & $0.10$ \\
		NGC~362  & $13$ & $1119$ & $17.94$ & $1.43$ & $0.11$ \\
		M5       &  $8$ &  $770$ & $18.12$ & $1.37$ & $0.12$ \\
		47~Tuc~E & $10$ & $1019$ & $16.87$ & $1.54$ & $0.16$ \\
		47~Tuc~W &  $8$ &  $749$ & $10.84$ & $1.58$ & $0.15$ \\
		\hline
	\end{tabular}
\end{table*}

\subsection{Photometric calibration}
\label{sec:phot_cal}

Instrumental CMDs for nine clusters (M12, NGC~6362, M4, M55, M22, NGC~6752, NGC~3201, M5 and 47~Tuc) 
were calibrated using linear transformations to already existing standard CMDs for these clusters 
\citep{mazur2003,kaluzny2013a,kaluzny2015a,kaluzny2015b}. The transformations followed:
\begin{equation}
\begin{aligned}
    v = V + a_{1} + a_{2}(B-V) \\
    b = B + b_{1} + b_{2}(B-V) \\
    b-v = c_{1} + c_{2}(B-V)
	\label{eq:bv}
\end{aligned}
\end{equation}
where $v$, $b$ and $b-v$ are instrumental and $V$, $B$ and $B-V$ are standard magnitudes and colors, 
respectively, and $a_1$, $a_2$, $b_1$, $b_2$, $c_1$ and $c_2$ are transformation coefficients. 
Calibration of M10 was based on $24$ Landolt standards \citep{landolt1992} observed within the CASE 
project with the Swope telescope and SITe\#3 camera in five Landolt fields on May $29^{th}$ $1998$. 
M30 and NGC~362 were calibrated on a base of $45$ standard stars from three Landolt fields observed 
on Aug $6^{th}$ $2000$. 

\section{Proper motions}
\label{sec:pm}

\subsection{Measurements}
\label{sec:pm_measurements}

The procedure employed to derive relative PMs was similar to that of \citet{anderson2006} 
\citep[also described in][]{zloczewski2011, zloczewski2012}, in which positions of stars in different 
epochs are determined with respect to nearby cluster members. This is the so called \emph{local 
transformation method}. Before attempting the measurements, we removed stars with relatively large 
magnitude errors ($\sigma_V$) from the lists of profile photometry. The following procedure was 
used: (i) a~curve of the form $f = a \cdot \exp(V) + b$ was fitted to magnitude--magnitude error 
relation, where $a$, $b$ are fitting parameters, (ii) for every subframe a~magnitude $V_S$ was 
selected such that for $V<V_S$ stars located above the curve were rejected, (iii) for $V>V_S$ stars 
with $\sigma_V>1.3f$ were rejected. Next, from the master list we selected candidates for 
\emph{grid stars} which were used as stable points for geometrical transformations between 
subframes. 

As a~first guess we selected stars located on main sequence (MS) and red giant branch (RGB) of the 
cluster's CMD. Those on the blue part of the horizontal branch (HB) had to be excluded because 
of  \textit{differential chromatic refraction} (DCR) which causes relative shifts in positions of stars 
of different colors \citep[e.g.][] {anderson2006}. To that end, a simple color criterion would be sufficient. 
However, CMDs of some clusters (e.g. M4, M22 or NGC 6362) contain appreciable numbers of stars located 
to the left of the RGB and above the subgiant branch (SGB), most of which must be field interlopers. 
In such cases, employing a color criterion would contaminate the first-guess grid sample. To avoid this, 
we applied the above population criterion instead, and, as a consequence, red HB stars were lost from 
the grid sample. However, since in all clusters they are a minute fraction ($\leq$0.03) of the MS/SGB/RGB 
population, excluding them had practically no effect on the results. 

The photometric quality of grid candidates was estimated based on the values of the CHI and 
SHARP parameters returned by ALLSTAR. As in \citet{zloczewski2012}, only stars with $0.02 \leq CHI \leq 1.00$ 
and $-0.3 \leq SHARP \leq 0.3$ were included. To each star, grid stars located in a~circle with a~radius 
of $150$~pixels ($\approx$1.09~arcmin) centered on that star were assigned. If there were fewer than 
$40$ stars inside the circle, the radius was successively increased by $50$~pixels ($\approx$0.36~arcmin) 
until this condition was fulfilled. The number of grid stars ranged from a~few tens to a~few hundreds per 
star. For each star, a~local geometrical transformation was found between the positions of grid stars on 
the reference frame and their positions on each single frame of a~given cluster. To that end, we used 
two--dimensional $3$rd order Chebyshev polynomials available in IRAF\footnote{IRAF (Image Reduction and 
Analysis Facility) is distributed by the National Optical Astronomy Observatories, operated by Association 
of Universities for Research in Astronomy (AURA), Inc., under cooperative agreement with the National 
Science Foundation (NSF).} tasks \emph{immatch.geomap} and \emph{immatch.geoxytran}. 

\begin{figure}
	\includegraphics[width=\columnwidth]{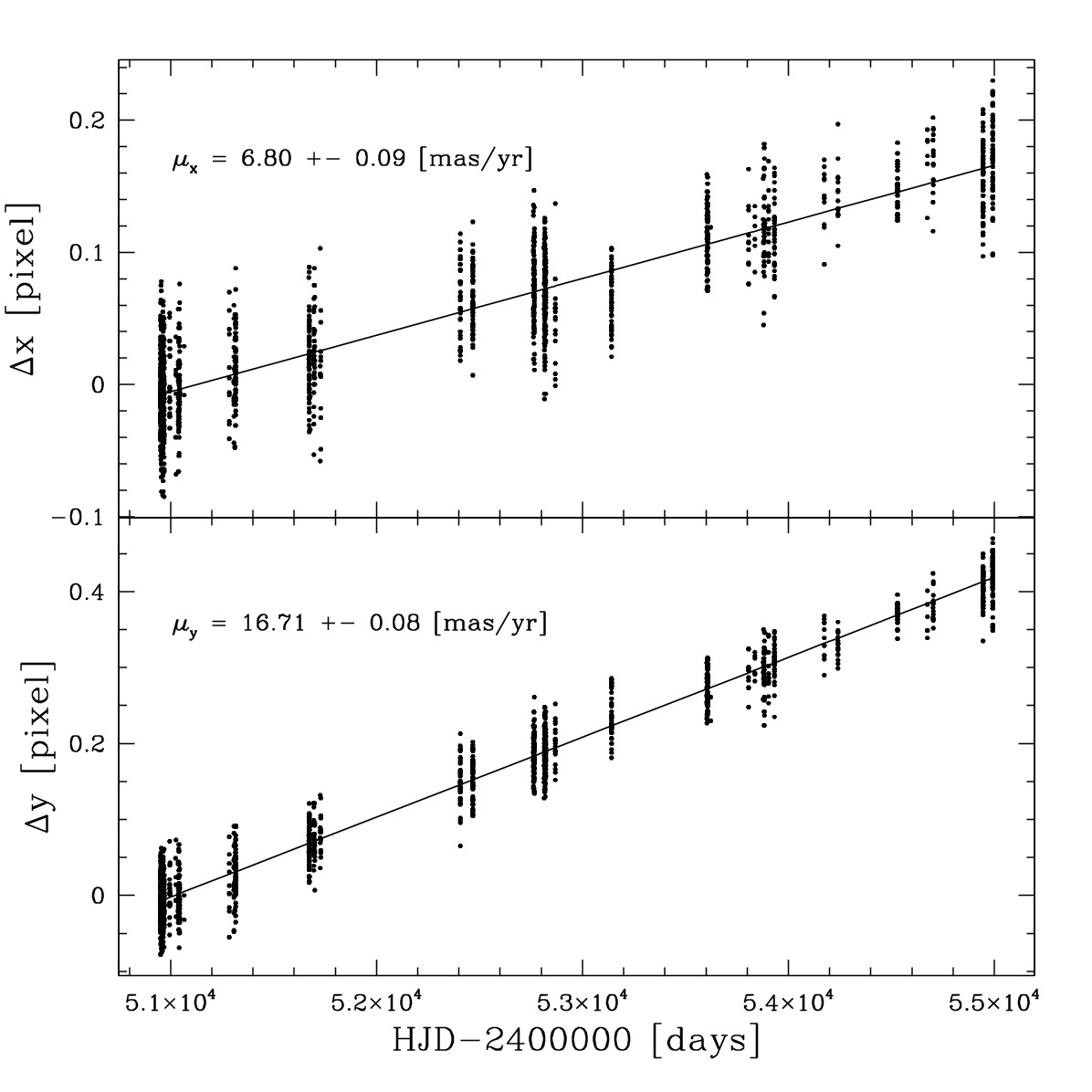}
    \caption{An example of the PM determination of star \#$150770$ from M4 with $V=18.149$~mag. 
    Proper motion is the slope of the linear fit to $\Delta x$ and $\Delta y$ versus time. 
    The $\Delta x$ and $\Delta y$ are differences between reference position of a~given star and 
    position measured at given epoch.}
    \label{fig:fit}
\end{figure}

Subsequently, coordinates ($X_{R}, Y_{R}$) of a~given star on the reference image were transformed 
into expected coordinates ($X_{C}, Y_{C}$) of this star, and then relative motions were derived 
as a~difference between expected and observed positions $\Delta X = X_{C} - X_{O}$ and 
$\Delta Y = Y_{C} - Y_{O}$. Finally, PMs $\mu_{X}$ and $\mu_{Y}$ were derived from weighted linear 
least--square fits to $\Delta X$ and $\Delta Y$ as functions of time. The weight of a~point was defined 
as the square root of the sum of squared uncertainties of the grid transformation (returned by IRAF 
task \emph{immatch.geomap}) and the PSF fitting \citep{kuijken2002}. The fitting was attempted for 
objects with positions measured in at least three epochs spanning at least three years except for stars 
in M30 (see Table~\ref{tab:frames}). Confidence level was set to $99\%$, i.e., only results with 
significance of the fit greater than that value were considered as reliable measurements. 
Fig.~\ref{fig:fit} shows an example of fitting for the star \#$150770$ from M4. 

PM calculations were conducted in three steps. First, we determined PMs of grid stars. Whenever 
it was necessary, stars with large PMs were rejected, and this step was repeated. That happend 
when fast moving field stars were significantly affecting the transformation between frames. 
In the next step we calculated PMs for all stars in the field of a~given cluster basing on the 
new grid stars. Finally, we again selected new grid candidates from all stars (keeping the 
conditions described above) and repeated the last step. Such a~procedure allowed to minimize 
the effect of grid star motions. 

Table~\ref{tab:pm} summarizes the number of stars for which PMs were obtained.

\begin{table}
	\centering
	\caption{Number of stars with measured PMs.}
	\label{tab:pm}
	\begin{tabular}{lc} 
		\hline
		Field & number \\
		\hline
		M12      & $19120$ \\
		NGC~6362 & $21851$ \\
		M4       & $24304$ \\
		M55      & $39566$ \\
		M22      & $87030$ \\
		NGC~6752 & $37454$ \\
		NGC~3201 & $32795$ \\
		M30      & $12638$ \\
		M10      & $27895$ \\
		NGC~362  & $23879$ \\
		M5       & $37042$ \\
		47~Tuc~E & $42506$ \\
		47~Tuc~W & $43161$ \\
		\hline
	\end{tabular}
\end{table}

Equatorial coordinates of the measured stars were derived from an astrometric solution based on stars 
with $V<17$~mag from the Fourth U.~S. Naval Observatory CCD Astrograph Catalog 
\citep[UCAC4,][]{zacharias2012}. The average residual of the solution ranged from $0.''13$ for M22 
to $0.''29$ for NGC~6752. Translation from $(\mu_x, \mu_y)$ to $(\mu_{\alpha}\cos\delta, \mu_{\delta})$ 
was based on the same astrometric solution.

\subsection{Error discussion}
\label{sec:error} 

PM errors given in the catalogue are statistical errors calculated from linear fitting. 
The errors depend on the following effects:
\begin{description}
 \item[1)] decreasing signal to noise ratio for fainter stars,
 \item[2)] location of the star on the image (stars located close to the edges of the frame have 
 smaller numbers of grid candidates and noticeably larger PM uncertainties), 
 \item[3)] frequency of blending (blends are more likely for stars located closer to the center 
of a cluster, and for clusters at larger heliocentric distances),
 \item[4)] uncertainties of transformations between images,
 \item[5)] number of epochs and total number of exposures used for calculations,
 \item[6)] time span between first and last epoch,
 \item[7)] DCR effect.
\end{description}

\begin{figure}
	\includegraphics[width=\columnwidth]{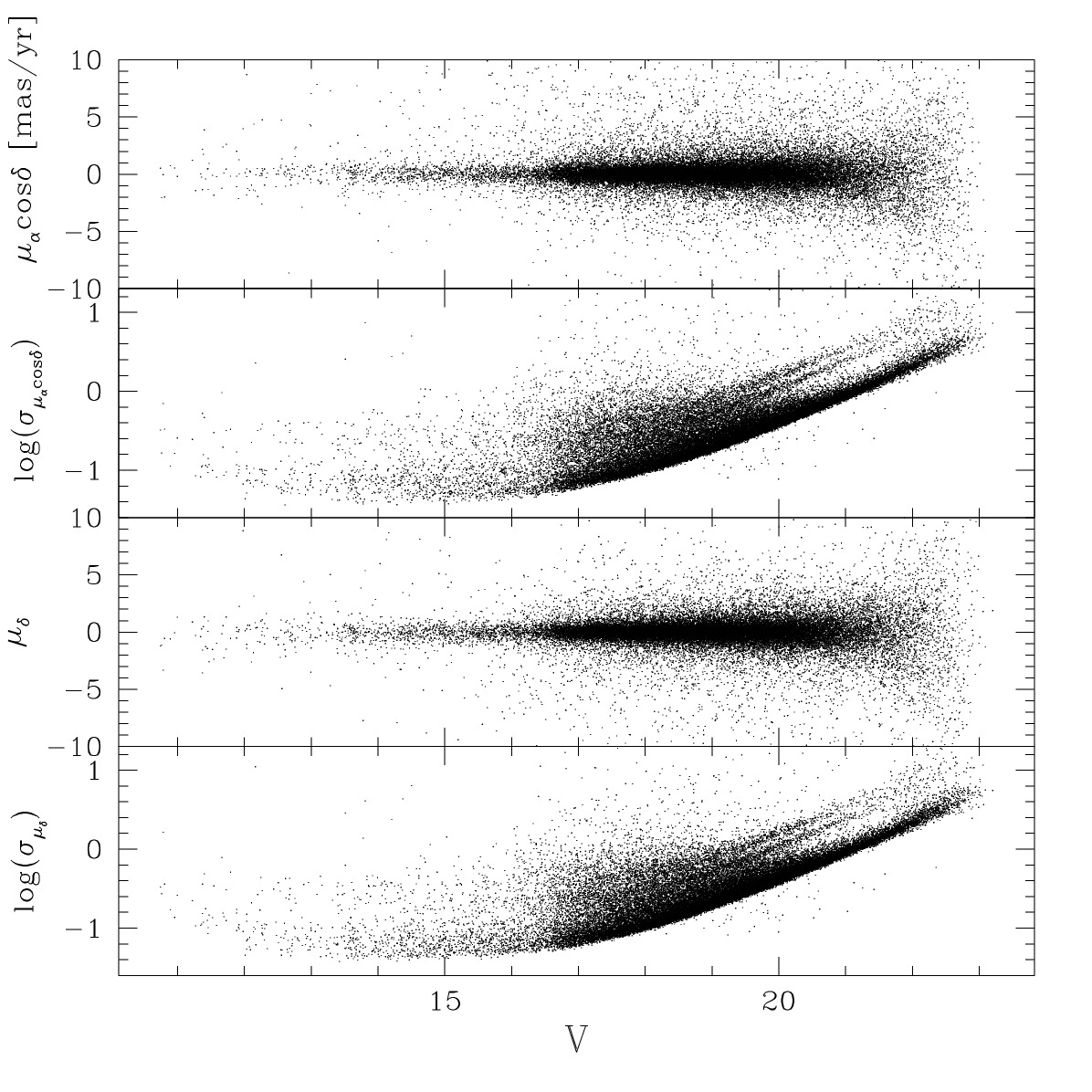}
    \caption{Plots of PMs and log of their uncertainties for NGC~6752.}
    \label{fig:ngc6752_err}
\end{figure}

The relevance of the first three effects is illustrated in Fig.~\ref{fig:ngc6752_err} for the 
example of NGC 6752. Characteristic is the growth of uncertainties for fainter stars, which is a~direct 
consequence of item 1). For red giants and horizontal branch stars the median of the uncertainties 
in both coordinates ranged from $0.04$~mas/yr in M4 to $0.25$~mas/yr in 47~Tuc W. For subgiants (SG) 
and stars located in the vicinity of the main sequence turn off point (MSTO) these values ranged from 
$0.05$~mas/yr in M4 to $0.38$~mas/yr in 47~Tuc W, and for stars from the lower MS - from $0.12$~mas/yr 
to $0.92$~mas/yr, respectively. 
The two weak branches visible in panels $2$ and $4$ of Fig.~\ref{fig:ngc6752_err} for $V>20$ mag are  
populated by stars from the edges of the reference image. The distance between these and the main branch 
is a measure of the combined inaccuracies related to item 2)  and to the fact that fewer 
epochs were available for these objects. Finally, points spread at the center of the plot ($15.5<V<19$)~mag 
are stars from the crowded center of the cluster, whose PM errors are mainly caused by blending.

The uncertainty of the transformations varies depending on star and epoch, and is a complex 
combination  of the employed formulae and the quality of the data to be transformed, which in turn depends 
on effects 1)--3). Its magnitude can be assessed from the vertical spread of points representing 
measurements in a given epoch (see Figure \ref{fig:fit} for an example).

Points 4) and 5) are discussed below for the example of 47~Tuc.
A~total of $2888$ stars were measured independently in the overlapping part of the E and W fields of 
this cluster, a~strip $300$~pixels ($2.175$~arcmin) wide. Note that we compare stars, first, located 
close to the edges of the subframes and second, from the vicinity of the cluster center. In both cases 
PM uncertainties are larger than elsewhere in the field. For this reason, we calculated the average 
PM differences in right ascension and declination for stars with PM uncertainties lower than 
$10$~mas/yr ($2712$ stars) only. The means were then $0.469 \pm 2.254$~mas/yr and $0.141 \pm 2.194$~mas/yr. 
After applying  $\sigma-clipping$ at the level of $3\sigma$ these values are equal to $0.374 \pm 2.03$~mas/yr 
and $0.179 \pm 2.0$~mas/yr. If we assume that PM uncertainties in the E and W fields can be expressed 
as $\frac{\epsilon}{\Delta t \sqrt{N}}$, where $\Delta t$ is the time base and $N$ is number of epochs 
used for PM determination in a~given field (see Table~\ref{tab:frames}), then by calculating the sum 
of their squares and assuming that it is equal to square of the uncertainty ($2$~mas/yr obtained above), 
we can obtain $\epsilon$. Then, by applying  the above formula with known $\epsilon$ the average PM 
uncertainties are $\approx$1.0~mas/yr and $\approx$1.7~mas/yr for the E and W fields, respectively. 
This means that the PM measurements in the E field are on average by $0.7$~mas/yr more accurate than 
in W. 
The difference of PM quality between these two fields comes mainly from field W. It has fewer epochs 
than for field E, and, even more importantly, the time span between the first and the last epoch was 
shorter by six years (see Table~\ref{tab:frames}).

Refraction affects the data in two ways. Firstly, by modifying (x,y) coordinates of stars 
with the same 
color by amounts depending on the air mass at the position of a given star. This effect is minimized 
by the local transformation method we apply in the present survey. Secondly, via the DCR effect 
mentioned above. The DCR increases the spread of points in a~single epoch, and may cause systematic 
shifts in PMs of stars with different colors. 
To assess its importance, we made linear fits to the relations between $\Delta x$ and $\Delta y$ and 
air mass for one of the bluest ($B-V=-0.028$ mag) and one of the reddest ($B-V=0.757$ mag) stars in 
NGC~6752 which has a relatively strong blue HB. For the red star, no significant deviations from zero 
were observed for the whole air mass range. For the blue star, at the largest air mass of 1.45 the deviation 
amounted to $\approx$0.06 px, which was significantly less than the $\Delta x$ or $\Delta y$ spread at 
any epoch. Moreover, about 80\% of the images were taken at air masses smaller than $\approx$1.3, at which 
the deviation was still at least two times smaller. We concluded that small systematic PM errors were 
likely for blue HB stars only, and did not attempt to introduce any corrections.

\begin{figure}
	\includegraphics[width=\columnwidth]{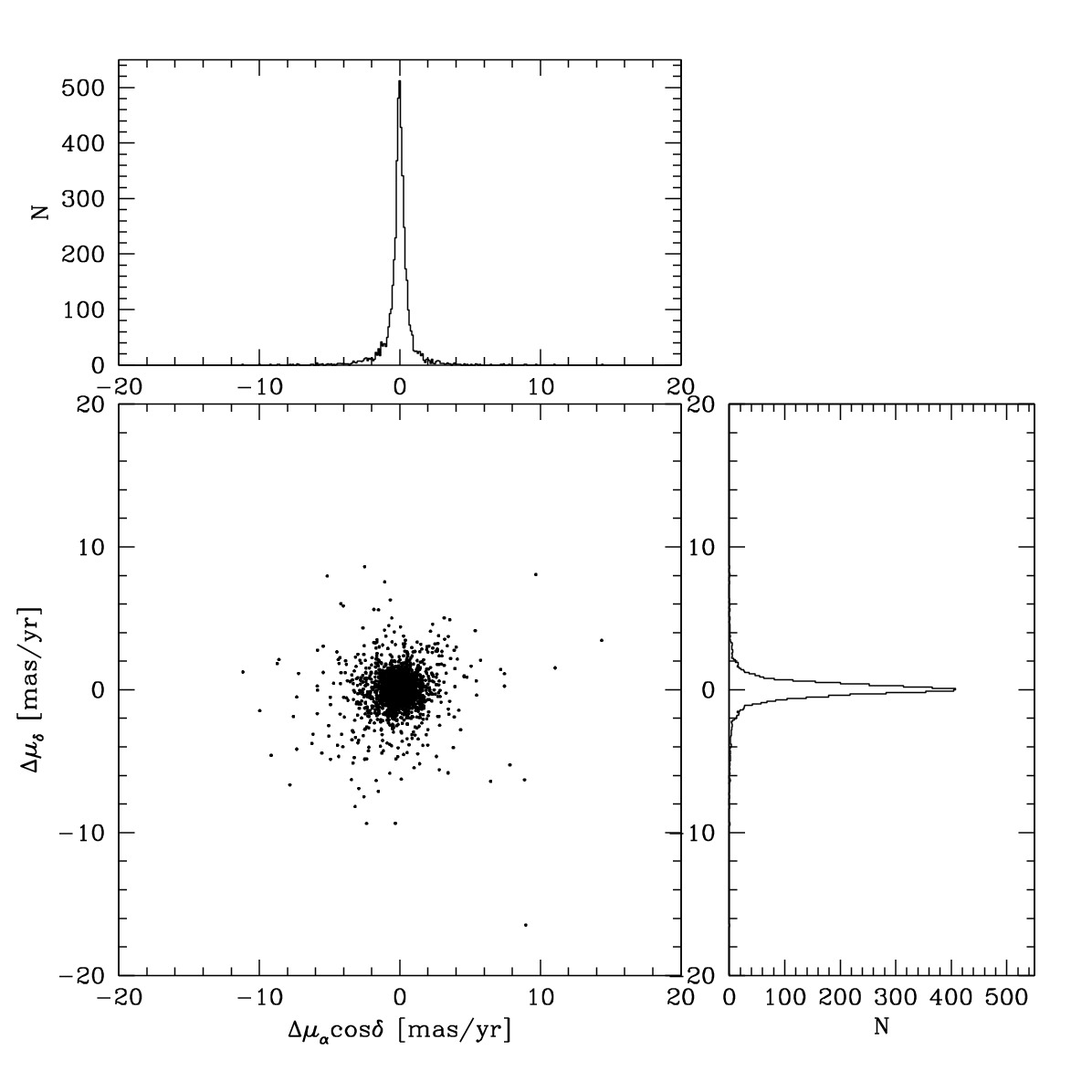}
    \caption{Differences between PMs of common stars from overlapping subframes in M4. 
    \emph{Panels}: VPD of the PM differences (\emph{lower left}); marginal histogram 
    of differences in right ascension, where $N$ are counts (\emph{upper left}); marginal 
    histogram of differences in declination (\emph{right}).}
    \label{fig:overlap}
\end{figure}

The overall consistency of the measurements was verified by comparing PMs of stars from overlapping 
parts of the subframes (where each star was measured independently at least two times). An example 
comparison is shown in Fig.~\ref{fig:overlap}. The symmetry of the Vector Point Diagram (VPD) indicates 
that the differences are random. Also, the marginal distributions take a Gaussian shape which further 
confirms this conclusion. Average PM differences depend on subframes taken for the comparison, but all 
of these are equal to zero within the errors. After rejecting outliers, the average differences 
$(\Delta \mu_{\alpha}\cos\delta, \Delta \mu_{\delta})$ between stars from overlapping regions of the 
four center subfields are $(-0.07 \pm 1.14, 0.06 \pm 1.16)$~mas/yr, between those of the twelve inner 
subfields (without the four center ones) are $(-0.05 \pm 0.77, 0.03 \pm 0.79)$~mas/yr and between those 
of the the remaining eight subfields are $(-0.28 \pm 1.67, 0.27 \pm 1.57)$~mas/yr.

\subsection{Comparison with other catalogues}
\label{sec:compare} 

\begin{figure}
	\includegraphics[width=\columnwidth]{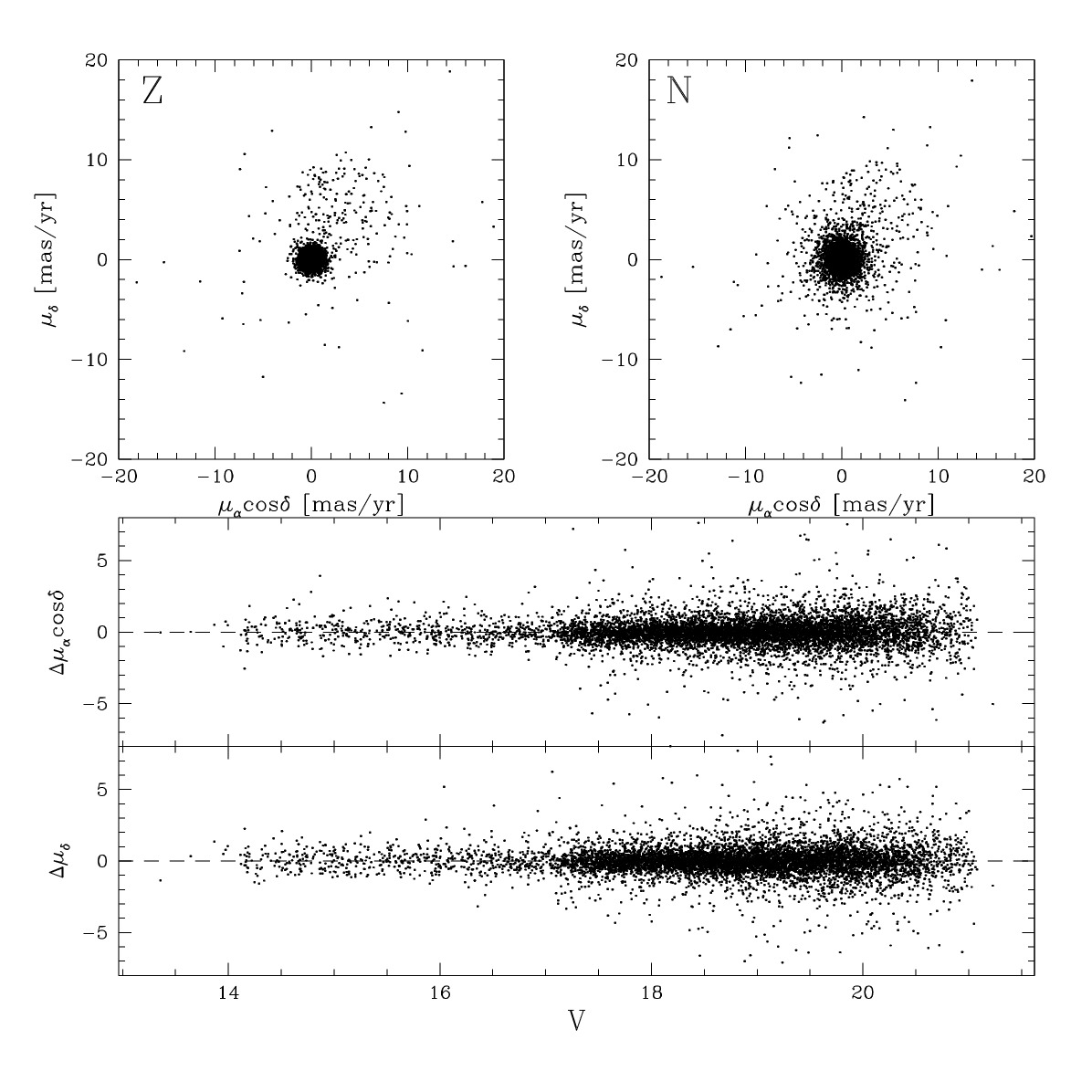}
    \caption{Comparison of PMs of common stars from \citet{zloczewski2011} (Z) and this work (N)
    for M55. \emph{Bottom panels}: PM differences in right ascension and declination.}
    \label{fig:kzcom}
\end{figure}

\begin{figure}
	\includegraphics[width=\columnwidth]{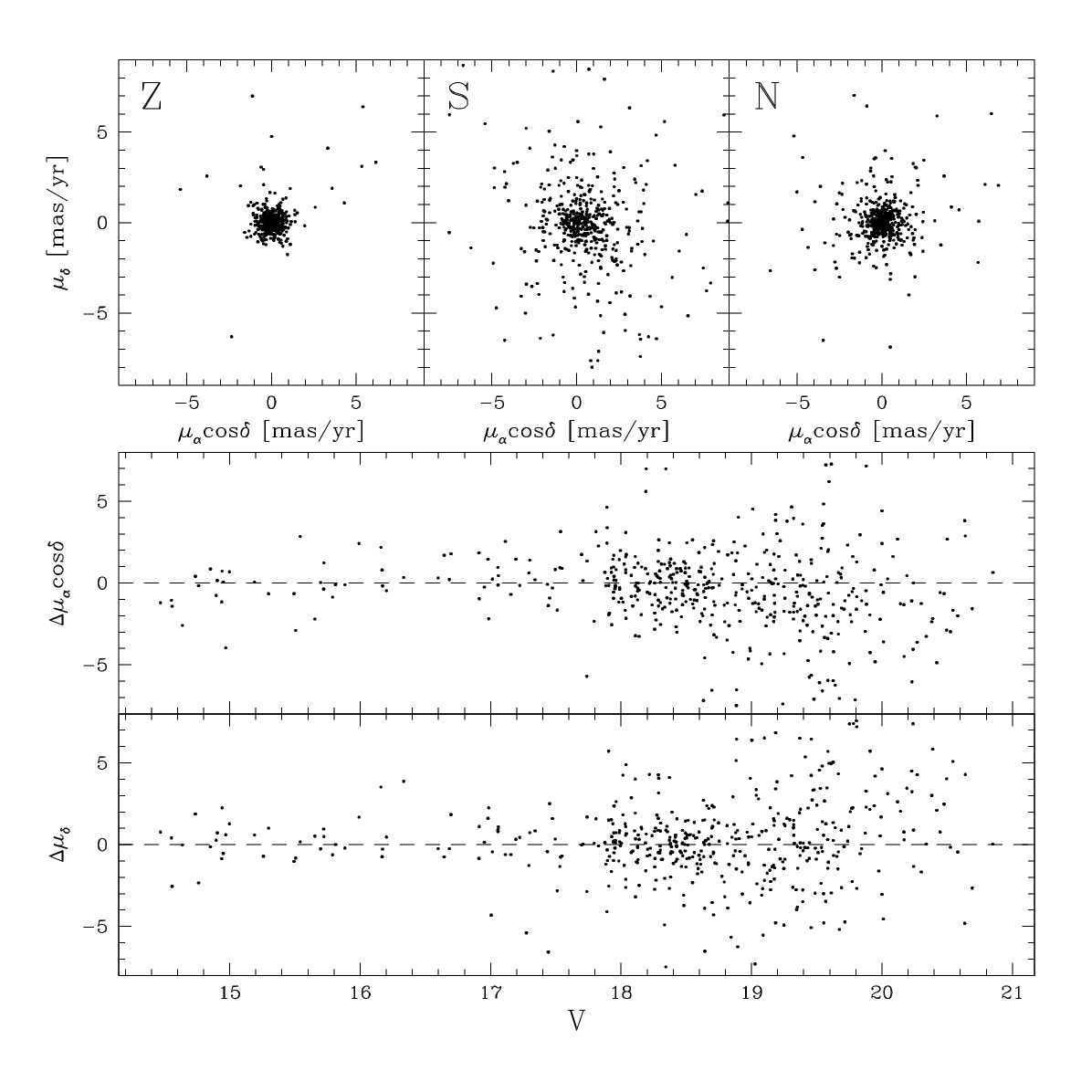}
    \caption{Comparison of PMs of common stars from \citet{zloczewski2011} (Z), \citet{sariya2012} 
    (S) and this work (N) for M55. \emph{Bottom panels}: PM differences in right ascension and 
    declination ($S-N$).}
    \label{fig:scom}
\end{figure}

An example comparison of our PMs with other catalogues is presented on Fig.~\ref{fig:kzcom} 
and~\ref{fig:scom} for the case of M55.

\citet{zloczewski2011} calculated PMs in M55 in four overlapping fields F1--F4. They used images 
at a~resolution of $0.259''$/pixel. Fig.~\ref{fig:kzcom} shows the comparison of PMs for $7822$ 
common stars in M55 from field F1 in their catalogue with those calculated in this work. F1 was 
chosen because of large number of common stars and long time base ($11$~years). The two top panels 
in Fig.~\ref{fig:kzcom} are the VPDs of common stars \citep[left~--][right~-- this work]{zloczewski2011}. 
As one could expect given the difference in telescope/camera resolution, the left plot has less spread 
than the right one. The average difference of PMs for all stars is $0.020 \pm 0.013$~mas/yr in right 
ascension and $-0.018 \pm 0.029$~mas/yr in declination. Both values are close to zero, and no systematic 
trends are visible (see the two bottom panels in Fig.~\ref{fig:kzcom}). The standard deviation is $1.16$ 
and $2.54$~mas/yr, respectively.

Fig.~\ref{fig:scom} presents the comparison of $480$ stars in common with the catalogues of 
\citet{zloczewski2011}, \citet{sariya2012}, and this work. \citet{sariya2012} used $26$ frames 
obtained at a~resolution of $0.238''$/pixel. The time span between the two epochs they only used 
was seven~years (five~years less than in the present work). The top panels in Fig.~\ref{fig:scom} 
are the VPDs of the common stars for \citet[][left]{zloczewski2011}, \citet[][middle]{sariya2012} 
and this work (right). The middle panel has significantly more spread than the two others. The 
average PMs difference between \citet{sariya2012} and this work for all common stars is 
$-0.511 \pm 0.160$~mas/yr in right ascension and $0.569 \pm 0.457$~mas/yr in declination, and the 
standard deviation is $3.510$ and $10.012$~mas/yr, respectively. The mean values are noticeably 
different from zero, futhermore the spread around them is large (see the two bottom panels on 
Fig.~\ref{fig:scom}). A longer time base and a larger number of exposures used for the PM measurements 
are  advantages of our results compared to those of \citet{sariya2012}.

\subsection{Completeness}
\label{sec:compl} 

\begin{figure}
  \centering
  \subfloat[M4]{\label{fig:m4_compl}
    \includegraphics[width=0.35\textwidth, angle=-90]{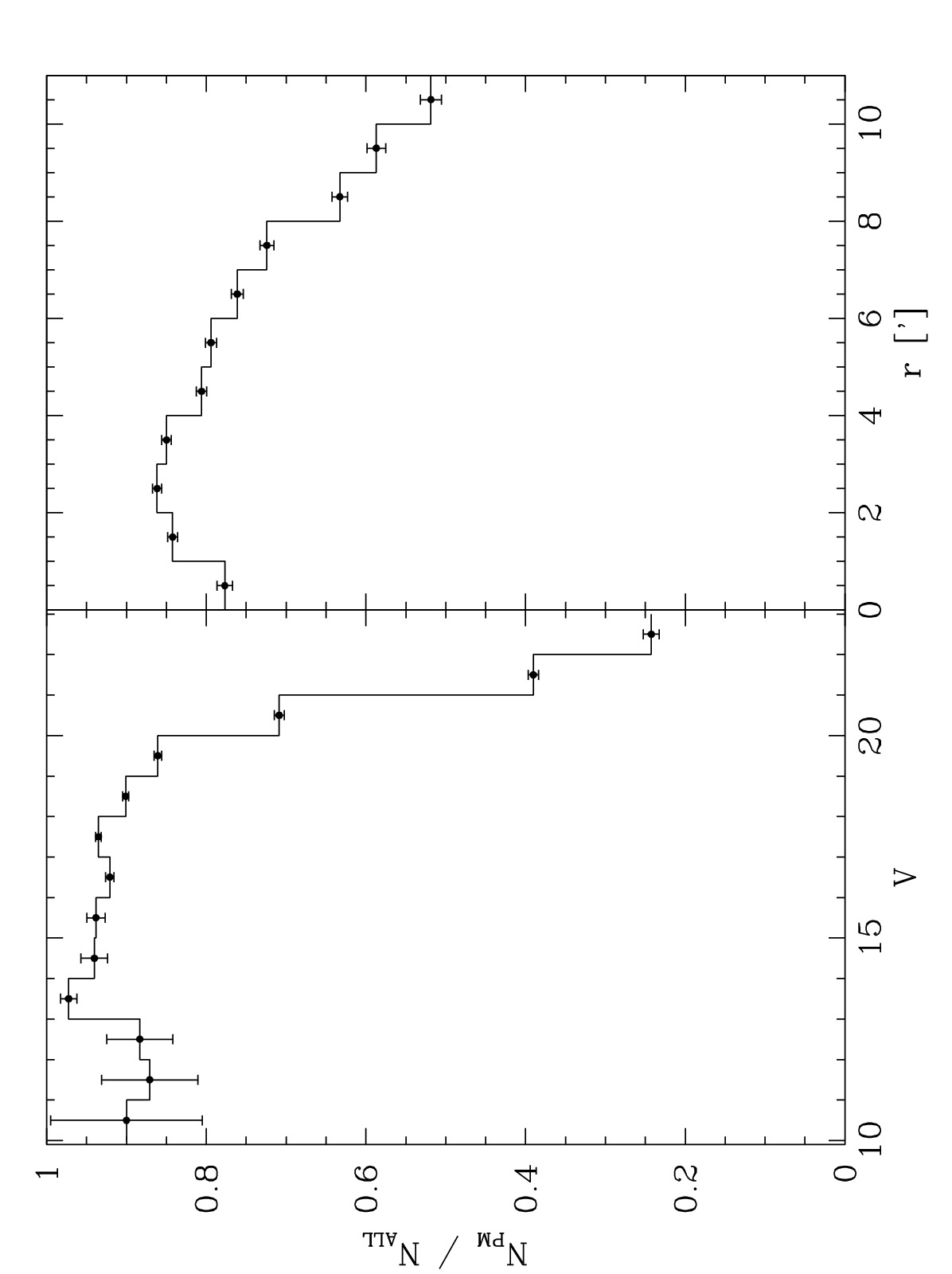}} \\
  \quad
  \subfloat[47~Tuc~W]{\label{fig:47TucW_vpd}
    \includegraphics[width=0.35\textwidth, angle=-90]{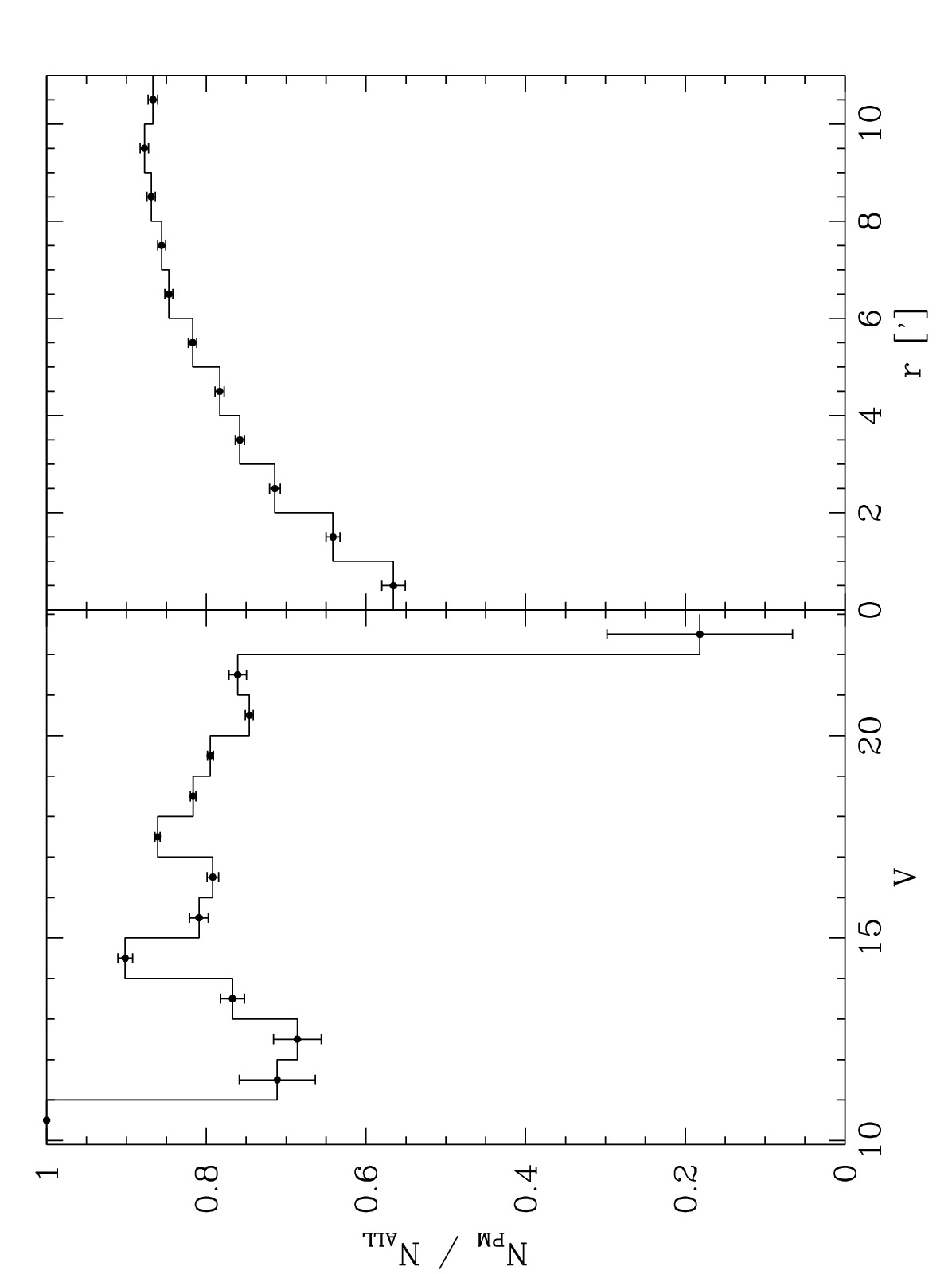}}
  \caption{Completeness of our PM catalogue for M4 (a) and 47~Tuc~W (b) defined as $N_{PM}~/~N_{ALL}$, 
  where $N_{PM}$~-- number of stars with PMs and $N_{ALL}$~-- number of master list stars in a~given 
  interval of $V$ or $r$. The uncertainties were estimated from the bimodal distribution. The $100\%$ 
  completeness in the $10-11$~mag range for 47~Tuc~W is misleading, because there is just one star 
  in it.}
\label{fig:2x_compl}
\end{figure}

For a~given range of magnitudes or angular radial distance from the cluster center ($r$), the 
completeness of the measurements was defined as the ratio of the number of stars for which PMs 
were successfully measured to the number of stars in the master lists in that range. The completeness 
was measured in intervals of $1$~mag in $V$ and $1$~arcmin in $r$. The results for the examples of M4 
and 47~Tuc~W are shown in Fig.~\ref{fig:2x_compl}. No attempt was made to estimate the completeness 
of the master lists.

In most GCs the completeness of PM determinations exceeded $70\%$ for $13 < V < 19$~mag, but dropped 
to $\approx$20\% for stars fainter than $20$~mag. As a~function of $r$, the completeness first 
increases (exceeding $70\%$ at $\approx$3$'$ ) but then  drops to $\approx$50\% or even a few percent 
in M30 and NGC~362. The minimum at small $r$ is due to the crowding at the centers of the clusters, 
and the decrease at large $r$ is caused by the fact that for more distant objects fewer epochs are 
available. In the case of 47~Tuc (Fig.~\ref{fig:47TucW_vpd}) we observe a~steady increase of the 
completeness with $r$ in both E and W field. This is because the center of the cluster is located 
at the edge of the reference frame.

\subsection{Cluster membership and membership probability}
\label{sec:memprob} 

For the purpose of the text clarity, VPDs for the investigated clusters are 
shown in the Appendix,. They illustrate different degrees of the separation between field and cluster 
stars. In all cases candidate cluster members concentrate at the coordinate origin. In M4, M22 and NGC~3201 
field stars form well defined clumps clearly separate from  the cluster stars. In the remaining cases, field 
and cluster stars overlap partly (e.g. M10) or entirely (e.g. M30). The clump at $(\mu_{\alpha}\cos\delta,
\mu_{\delta})=(0.10 \pm 0.01, 7.44 \pm 0.01)$~mas/yr in the VPD of M55 is composed of stars from the 
Sagittarius dwarf galaxy. Similar clumps in VPDs of NGC~362 at $(\mu_{\alpha}\cos\delta,\mu_{\delta})=(-5.76 
\pm 0.01, 1.45 \pm 0.01)$~mas/yr and 47~Tuc at $(\mu_{\alpha}\cos\delta,\mu_{\delta})=(-4.54 \pm 0.02, 1.09 
\pm 0.01)$~mas/yr contain members of the SMC.

For each star with a derived PM the \emph{membership status} was obtained and the \emph{membership 
probability} was calculated. To estimate membership status, we followed the method described by 
\citet{zloczewski2011, zloczewski2012}. Based on the location of a~star on the VPD and its PM error 
($\sigma_{\mu}$) we assigned it to one of the three membership classes; $mem=0,1$ or $2$ for 
non--members, probable members and members of the cluster, respectively. To that end, we first divided 
stars into magnitude bins, each containing $100$ stars. For every bin we calculated average values and 
standard deviations of $\mu_{\alpha}\cos\delta$ and $\mu_{\delta}$ ($M_{\alpha}$, $M_{\delta}$, 
$S_{\alpha}$, $S_{\delta}$), and PM errors ($ME_{\alpha}$, $ME_{\delta}$, $SE_{\alpha}$, $SE_{\delta}$). 
Next, the parameters $S = (S_{\alpha}^2 + S_{\delta}^2)^{1/2}$, $ME = (ME_{\alpha}^2 + ME_{\delta}^2)^{1/2}$ 
and $SE = (SE_{\alpha}^2 + SE_{\delta}^2)^{1/2}$ were determined. Stars with $\mu > 3 \cdot S$ were 
considered non--members ($mem=0$), stars with $\mu \leq 3 \cdot S$ but $\sigma_{\mu} > ME + 3 \cdot SE$ 
were classified as probable cluster members ($mem=1$), and finally stars with $\mu \leq 3 \cdot S$ and 
$\sigma_{\mu} \leq ME + 3 \cdot SE$ were classified as cluster members ($mem=2$). Table~\ref{tab:status} 
lists the numbers of stars assigned to each membership class in the analyzed GCs.

\begin{table}
	\centering
	\caption{Statistics of the membership status of stars in the analyzed GCs.}
	\label{tab:status}
	\begin{tabular}{lrrr} 
		\hline
		Cluster & $mem=0$ & $mem=1$ & $mem=2$ \\
		\hline
		M12      &  $3050$ &  $435$ & $15635$ \\
		NGC~6362 &  $4913$ &  $302$ & $16636$ \\
		M4       &  $5870$ &  $529$ & $17905$ \\
		M55      &  $7058$ &  $772$ & $31736$ \\
		M22      & $45481$ & $1353$ & $40196$ \\
		NGC~6752 &  $4106$ &  $742$ & $32606$ \\
		NGC~3201 &  $6911$ &  $562$ & $25322$ \\
		M30      &  $1816$ &  $312$ & $10510$ \\
		M10      &  $3138$ &  $620$ & $24137$ \\
		NGC~362  &  $6311$ &  $548$ & $17020$ \\
		M5       &  $6418$ &  $992$ & $29632$ \\
		47~Tuc~E &  $4702$ &  $924$ & $36880$ \\
		47~Tuc~W &  $5623$ &  $975$ & $36563$ \\
		\hline
	\end{tabular}
\end{table}

The first attempt to estimate cluster membership probability ($P_{\mu}$) was undertaken by 
\citet{vasilevskis1958}, who proposed the formula  
\begin{equation}
    P =  \frac{\Phi_c(\mu_x, \mu_y)}{\Phi_c(\mu_x, \mu_y) + \Phi_f(\mu_x, \mu_y)},
	\label{eq:P}
\end{equation}
where $\Phi_c(\mu_x, \mu_y)$ and $\Phi_f(\mu_x, \mu_y)$ are bivariate gaussian functions describing 
VPD distributions of cluster and field stars, respectively. \citet{jones1988} pointed out that 
probabilities resulting from Eq.~(\ref{eq:P}) are biased due to two effects. First, parameters in 
both distributions depend on the brightness of stars. Second, as opposed to the field stars, the 
spatial distribution of cluster members is not uniform: their number density decreases with increasing 
$r$. Disregarding these two facts leads to underestimation and overestimation of the probabilities 
for bright and faint stars, respectively. The same holds, respectively, for stars located at the 
center of the cluster and at its outskirts. \citet{jones1988} introduced two important modifications. 
To eliminate the problem of magnitude dependence, they  fitted bivariate functions in overlapping magnitude 
bins. To deal with the second problem, they adopted a~model in which the dependence of the surface density 
of stars on distance from the cluster center was taken into account. A~modification of the above method 
is the \emph{local sample method} described in detail e.g. by \citet{kozhurina-platais1995}, where for 
each target star functions $\Phi_c$ and $\Phi_f$ are fitted to stars from magnitude and distance ranges 
centered on that star. Another significant modification was introduced by \citet{girard1989}, who proposed 
to smooth each point on the VPD by replacing it with a~normal distribution with $\sigma$ equal to the 
PM error. This way a~continuous distribution was obtained, to which the bivariate functions were fitted. 
In this work, we followed the above methods, slightly modifying them as described below. 

First, each $i$--th star on the VPD was replaced with a~two-dimensional normal distribution. 
Next, a~dense grid of ($\mu_x$, $\mu_y$) points was defined on the VPD, and at each grid point 
the function:
\begin{equation}
    f(\mu_x, \mu_y) = \displaystyle\sum_{i=1}^{N} \displaystyle\sum_{j=1}^{N} 
    \frac{1}{2\pi\varepsilon_{x_i}\varepsilon_{y_i}} 
    \exp\left[-\left(\frac{(\mu_x - \mu_{x_i})^2}{2\varepsilon_{x_i}^2} + 
    \frac{(\mu_y - \mu_{y_i})^2}{2\varepsilon_{y_i}^2}\right)\right] 
	\label{eq:ker}
\end{equation}
was evaluated, where ($\mu_{x_i}$, $\mu_{y_i}$) are PM components of the $i$th star, and 
($\varepsilon_{x_i}$, $\varepsilon_{y_i}$) are PM uncertainties. For all GCs the distance between 
grid points was $0.1$~mas/yr in both coordinates. The smooth distribution thus obtained was then 
approximated with two two--dimensional functions: a~circular Gaussian representing the cluster 
distribution
\begin{equation}
    \Phi_{c}(\mu_x, \mu_y) = A_1\cdot \exp\left[-\left(\frac{(\mu_x-\mu_{x01})^2}{2\sigma_x^2} 
    + \frac{(\mu_y-\mu_{y01})^2}{2\sigma_y^2} \right) \right]
	\label{eq:g1}
\end{equation}
and an~elliptical Gaussian representing field distribution:
\begin{align}
    \Phi_{f}(\mu_x, \mu_y) &= A_2\cdot \exp\Big[-\Big(a(\mu_x-\mu_{x02})^2 -  \nonumber \\ 
    &b(\mu_x-\mu_{x02})(\mu_y-\mu_{y02}) + c(\mu_y-\mu_{y02})^2 \Big) \Big]
	\label{eq:g2}
\end{align}
Both functions were fitted in circular apertures centered on $(\mu_{x01},\mu_{y01})$ and 
$(\mu_{x02},\mu_{y02})$ as a~first guess. Depending on the nature of the distribution of stars 
on the VPD, the apertures used to fit the Gaussians were \emph{separated} (for M4, M22 and NGC~3201) 
or \emph{concentric} (for the remaining GCs). The aperture radii were multiples of $\sigma$ defined 
as an aritmetic mean of $\sigma_x$ and $\sigma_y$. In total, eleven parameters were searched for: 
$A_1$, $A_2$, $\mu_{x01}$, $\mu_{y01}$, $\mu_{x02}$, $\mu_{y02}$, $\sigma_x$, $\sigma_y$, $a$, $b$, 
$c$. PM errors strongly depend on the magnitude of the star. To minimize this effect the fitting 
was performed in magnitude bins. This approach significantly reduces the computation time in comparison 
with the local sample method as originally proposed by \citet{kozhurina-platais1995}. In this study, 
bins $2$~mag wide in $V$ were used, with the upper limit of the first bin located above the horizontal 
branch, and the lower limit of the last bin at about $V=23$~mag (depending on the cluster). The bins 
were overlapping in such a~way that the upper limit of the next bin was shifted upwards by $0.025$ or 
$0.05$~mag with respect to the lower limit of the previous one (see Table~\ref{tab:par}). 
Table~\ref{tab:par} sumarizes the adopted parameters of apertures and bins.

\begin{table}
	\centering
	\caption{Parameters of apertures and bins adopted for Gaussian fitting: \emph{c}~-- 
	concentric apertures, \emph{s}~-- separated apertures, $r_{GC}$, $r_{F}$~-- fitting 
	radii for cluster and field, respectively, \emph{MRAN}~-- magnitude range in mag, 
	\emph{DBIN}~-- shift between magnitude bins in mag.}
	\label{tab:par}
	\begin{tabular}{lccccc} 
		\hline
		Cluster & aperture & $r_{GC}$ & $r_F$ & MRAN & DBIN \\
		\hline
		M12      & c & $3\sigma$ & $10\sigma$ & $13.0 - 23.5$ & $0.025$ \\ 
		NGC~6362 & c & $3\sigma$ & $10\sigma$ & $14.5 - 24.0$ & $0.025$ \\ 
		M4       & s & $3\sigma$ & $8\sigma$  & $13.0 - 21.2$ & $0.050$ \\ 
		M55      & c & $3\sigma$ & $15\sigma$ & $14.0 - 24.0$ & $0.025$ \\ 
		M22      & s & $3\sigma$ & $8\sigma$  & $13.0 - 23.5$ & $0.050$ \\ 
		NGC~6752 & c & $3\sigma$ & $12\sigma$ & $13.0 - 23.0$ & $0.025$ \\ 
		NGC~3201 & s & $3\sigma$ & $8\sigma$  & $13.0 - 24.0$ & $0.025$ \\ 
		M30      & c & $3\sigma$ & $12\sigma$ & $14.0 - 23.2$ & $0.025$ \\ 
		M10      & c & $3\sigma$ & $15\sigma$ & $13.5 - 23.0$ & $0.025$ \\ 
		NGC~362  & c & $3\sigma$ & $15\sigma$ & $14.0 - 23.5$ & $0.025$ \\ 
		M5       & c & $3\sigma$ & $15\sigma$ & $13.7 - 23.5$ & $0.025$ \\
		47~Tuc~W & c & $3\sigma$ & $15\sigma$ & $13.0 - 22.3$ & $0.025$ \\
		47~Tuc~E & c & $3\sigma$ & $15\sigma$ & $13.0 - 22.3$ & $0.025$ \\
		\hline
	\end{tabular}
\end{table}

For each star $\Phi_c$ and $\Phi_f$ were fitted using the magnitude bin whose center was closest 
to that star. For the brightest and faintest stars the first and last bins were used, respectively. 
In principle, the magnitude bins should be subdivided into distance bins, however due to the relatively 
small angular size of our field of view the spatial distribution was ignored. 

\begin{figure}
  \centering
  \subfloat[M4]{\label{fig:m4_prob}
    \includegraphics[width=0.35\textwidth, angle=-90]{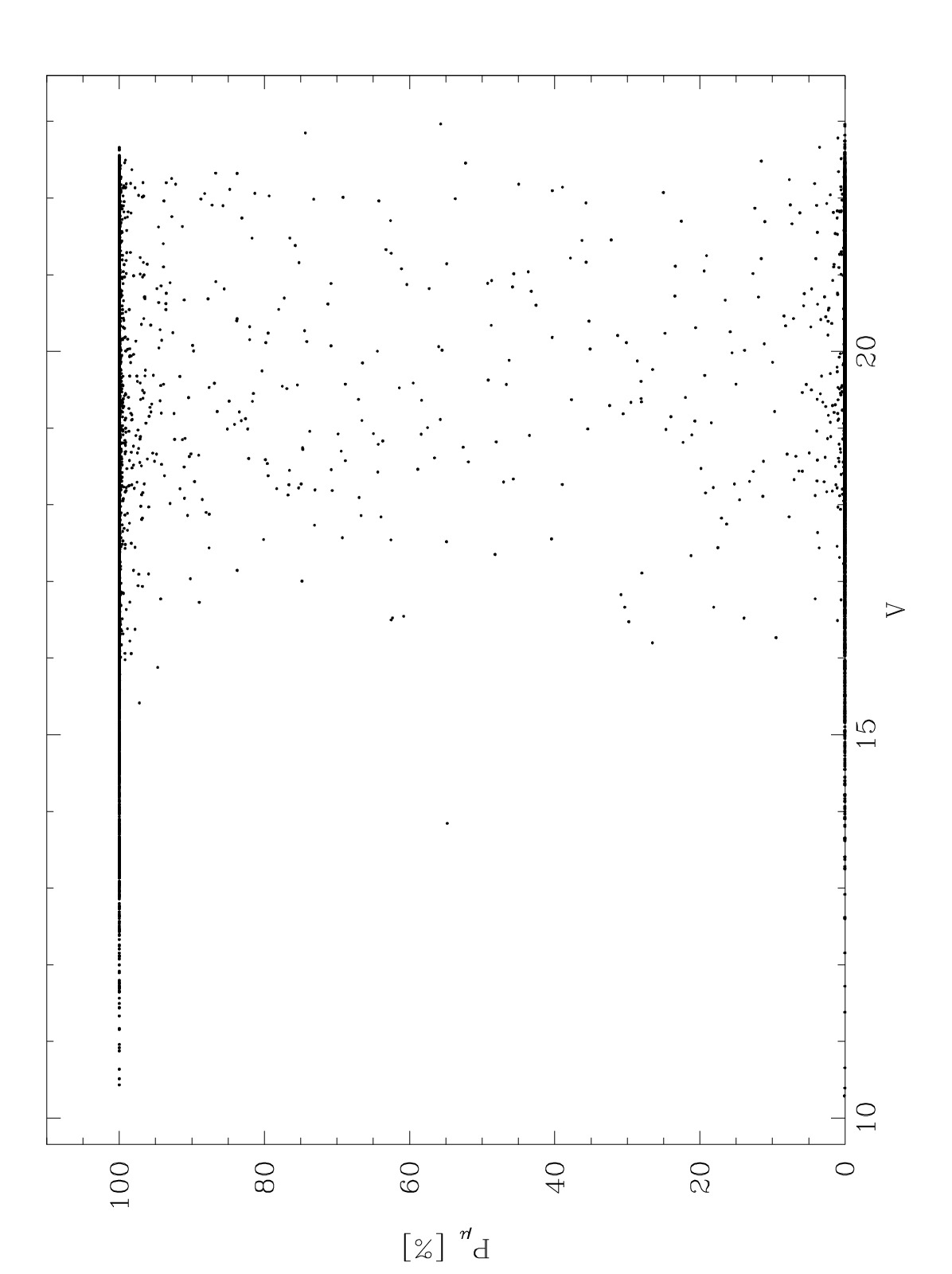}} \\
  \quad
  \subfloat[M55]{\label{fig:m55_prob}
    \includegraphics[width=0.35\textwidth, angle=-90]{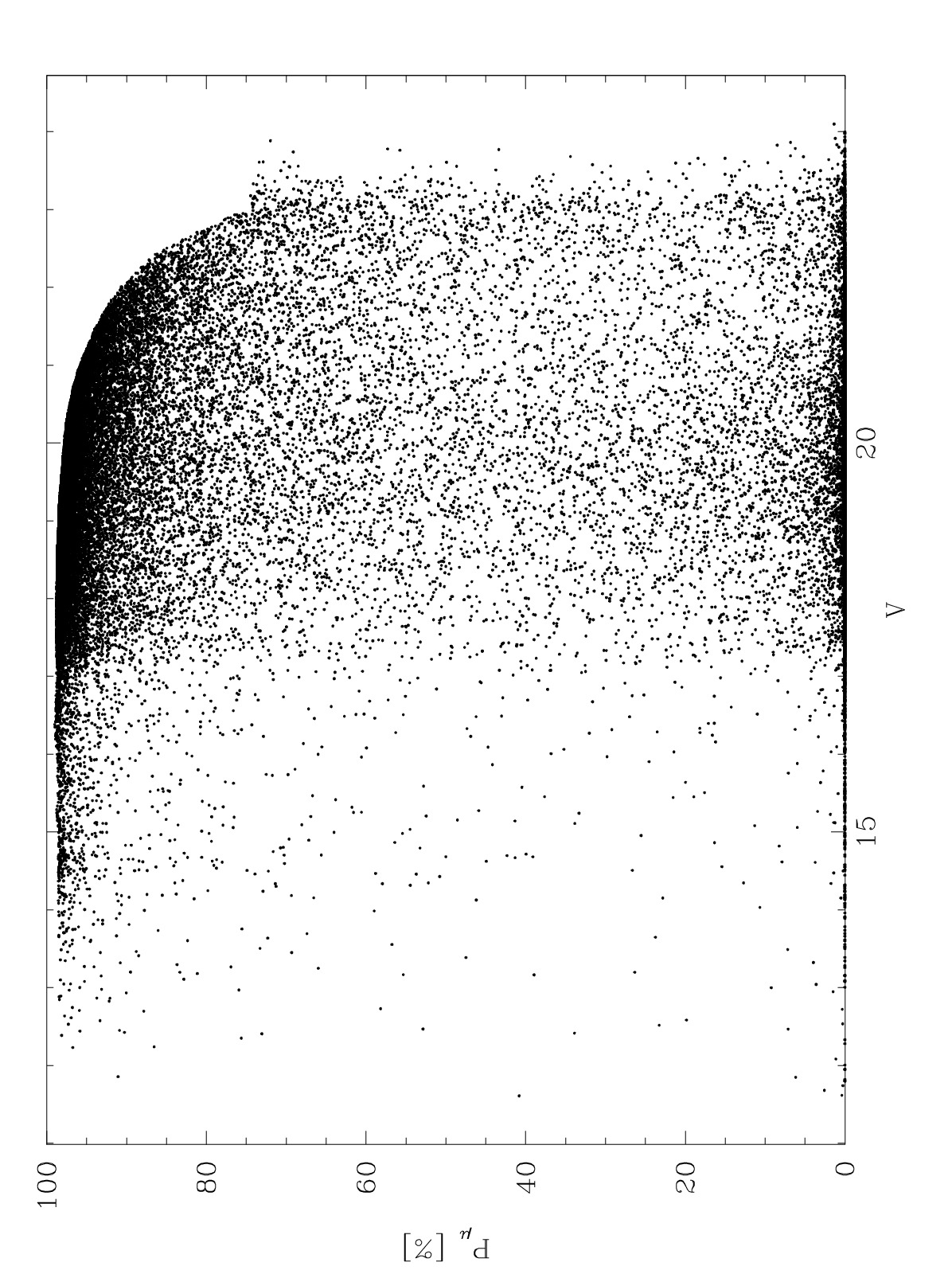}}
  \caption{Membership probabilities ($P_{\mu}$) as a~function of $V$ magnitude in well separated 
  (M4, see Fig.~\ref{fig:m4_prob}) and poorly separated GC (M55, see Fig.~\ref{fig:m55_prob}).}
\label{fig:2x_prob}
\end{figure}

In the end, membership probabilities were calculated using Equation~(\ref{eq:P}) with the improved 
methods of calculating $\Phi_c$ and $\Phi_f$ described above. Example results for well (M4) and poorly 
separated GC (M55) are presented in Fig.~\ref{fig:2x_prob}.

\subsection{Comparison of membership probabilities with catalogues of radial velocities}
\label{sec:radvel} 

\begin{figure*}
	\includegraphics[width=\textwidth]{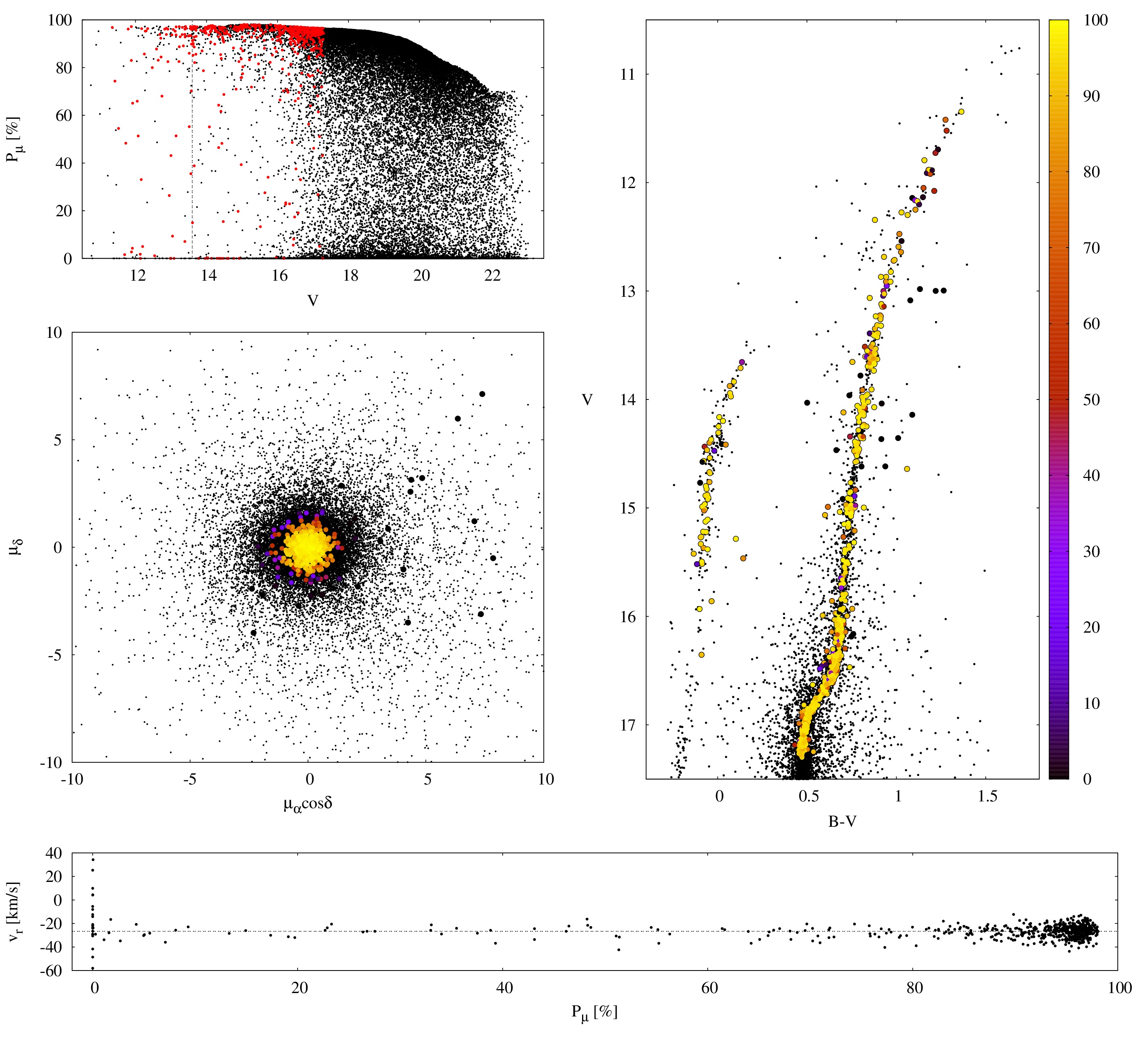}
    \caption{Globular cluster NGC~6752. \emph{Left top}: membership probability ($P_{\mu}$) 
    as a~function of $V$ magnitude. Stars for which radial velocities ($v_r$) are known are marked 
    with red points. The black vertical line indicates CCD saturation level. \emph{Middle left}: 
    VPD (central part of Fig.~\ref{fig:ngc6752vpd}). Stars with known $v_r$ are marked with larger 
    dots, $P_{\mu}$ is color coded. \emph{Right}: CMD for red giants, subgiants, horizontal branch 
    stars and stars from the MSTO vincinity. \emph{Bottom}: $v_r$ as function of $P_{\mu}$. Four 
    stars with $v_r$ larger than the axis scale are omitted.}
    \label{fig:ngc6752_radvel2}
\end{figure*}

We verified the reliability of the calculated membership probabilities in M4, NGC~3201, M55 and 
NGC~6752 by comparing the values obtained for individual stars brighter than the MSTO with their 
radial velocities ($v_r$) taken from appropriate catalogues. The example result for NGC~6752 is 
shown in Fig.~\ref{fig:ngc6752_radvel2}. In this case the analysis was done for $642$ red giants, 
subgiants, horizontal branch stars, and stars from the MSTO vincinity taken from the catalogue of 
\citet{lardo2014}. Radial velocities of those stars ranged from -88 to $105$~km/s. Only stars with 
velocities close to the mean heliocentric radial velocity ($v_h$) of NGC~6752 ($-26.7 \pm 0.2$~km/s 
\citep{harris1996}), $v_h \pm 3\sigma_v$ (where $\sigma_v = 4.9$~km/s), are potential cluster members, 
the remaining ones most probably do not belong to the cluster. Membership probabilities confirm this 
conclusion. The bottom panel of Fig.~\ref{fig:ngc6752_radvel2} shows radial velocities as a~function 
of the membership probability. Stars with high $v_r$ have low probabilities, and most stars with 
$v_r$ close to $v_h$ have high probabilities. On the top left panel these are marked by red points. 
The black vertical line indicates CCD saturation level. The middle left panel presents the central 
part of Fig.~\ref{fig:ngc6752vpd}, with color coded membership probabilities. Stars with high $P_{\mu}$ 
concetrate around the $(0,0)$ point while stars with low $P_{\mu}$ are located far from this point. 
On the right panel the same stars are marked on the CMD. 

For all four GCs considered in this subsection the overall result of the comparison shows a~very 
good consistency of membership probabilities (as well as membership status) with radial velocities. 
The majority of stars with $v_r$ close to the $v_h$ of a~given cluster have $mem=2$ and high membership 
probabilities (from $\overline{P_{\mu}}>80\%$ and median value $>90\%$ for M55 and NGC~6752 to 
$\overline{P_{\mu}}=100\%$ for M4 and NGC~3201, where $\overline{P_{\mu}}$ is the average membership 
probability of stars with known $v_r$). The only exceptions are most probably overexposed stars or 
stars whose PMs happen to be similar to the PM of the cluster. Stars with $v_r$ far from $v_h$ have 
$mem=0$ and $P_{\mu}<50\%$. 

This experiment allows to define the probability limit $P1$ such that objects with $P \geq P1$ will 
be considered as cluster members, and those with $P<P1$ as field stars. This limit will be subsequently 
used to clean the CMDs from interlopers. Keeping in mind that stars used in the above analysis are 
bright, we chose a~value slightly lower than $\overline{P_{\mu}}$ to make it appropriate also for 
fainter stars. The average $\overline{P_{\mu}} - \sigma_{\overline{P_{\mu}}}$, where 
$\sigma_{\overline{P_{\mu}}}$ is a~standard deviation, for stars with radial velocities close 
to $v_h$ from all four clusters, is equal to $69.3\%$. We adopt a~value of $P1=70\%$ for all GCs. The 
consistency of the probability-based cleaning procedure with a~given $P1$ can be assessed by determining 
how many stars with membership class $0$ or $1$ remain in the cleaned CMD, and how many stars with 
membership $1$ or $2$ are removed from the CMD. For the well separated clusters M4, M22 and NGC~3201 
the corresponding percentage points are $13.1$ and $3.5$, while for those poorly separated the values 
are $3.9$ and $33.5$, respectively. These cleaning procedures based on both membership status and 
membership probability are magnitude--dependent, and within each magnitude interval ($V-\Delta V, 
V+\Delta V$) they reduce to selecting stars from within a circle of a~radius $r_{\mu}$ centered on 
the $(0,0)$ point of the VPD. For status--based cleaning $r_{\mu}$ increases faster with magnitude 
than in probability--based cleaning. As~a result, the number of faint stars with $mem=2$ and 
$P_{\mu}<70\%$ can be significantly larger than the number of those with $mem=0$ and $P_{\mu} \geq 70\%$. 
This effect is most pronounced in poorly separated clusters.

\section{Color--Magnitude Diagrams}\label{sec:cmd}

\begin{figure*}
	\includegraphics[width=0.55\textwidth, angle=-90]{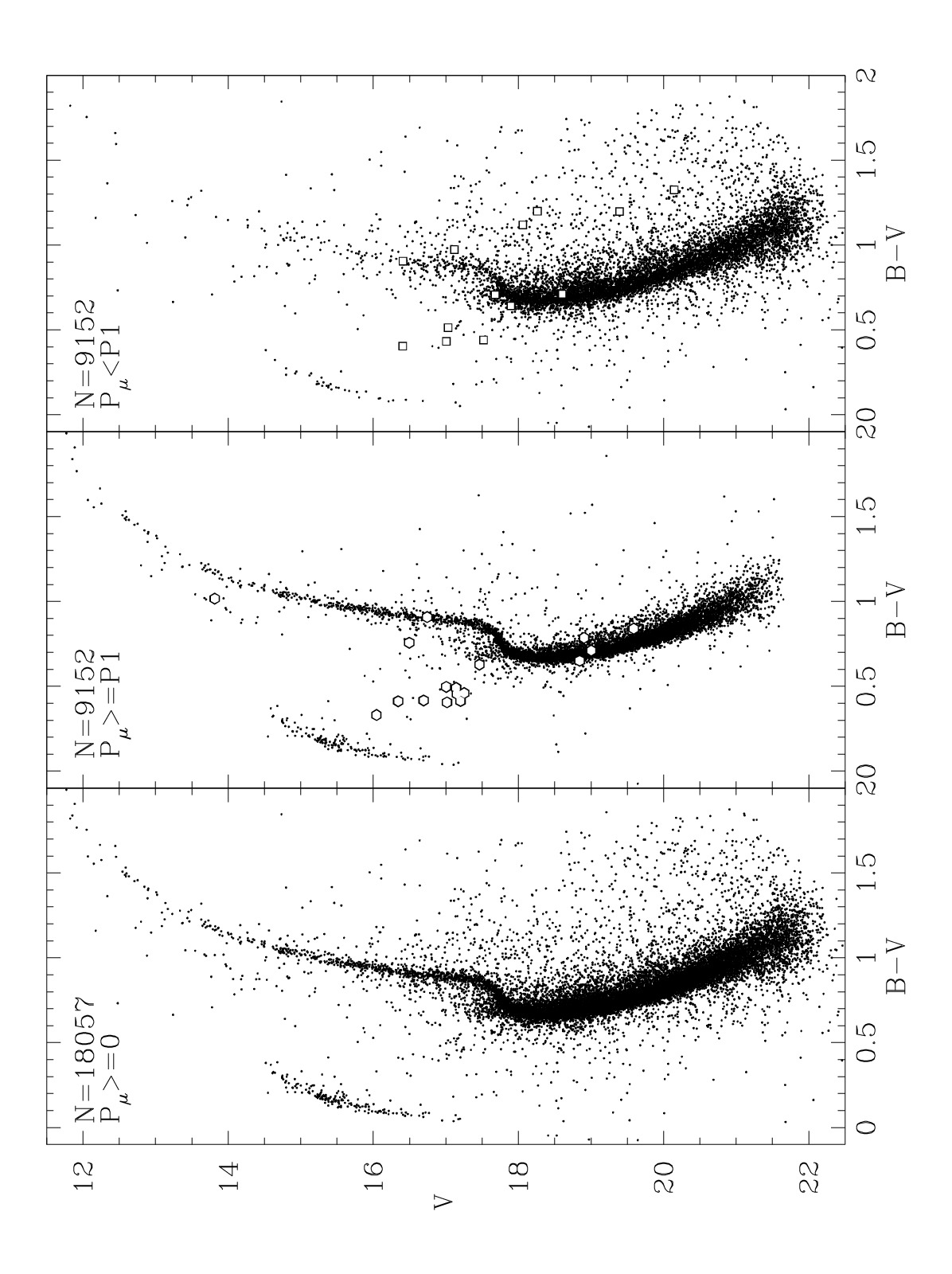}
    \caption{CMD of M12. All stars with PMs and colors measured (\emph{left}); stars with probabilities 
    $P_{\mu} \geq P1$ (\emph{middle}); stars with probabilities $P_{\mu} < P1$, where $P1=70\%$ 
    (\emph{right}).}
    \label{fig:m12_cmd}
\end{figure*}

\begin{figure*}
	\includegraphics[width=0.55\textwidth, angle=-90]{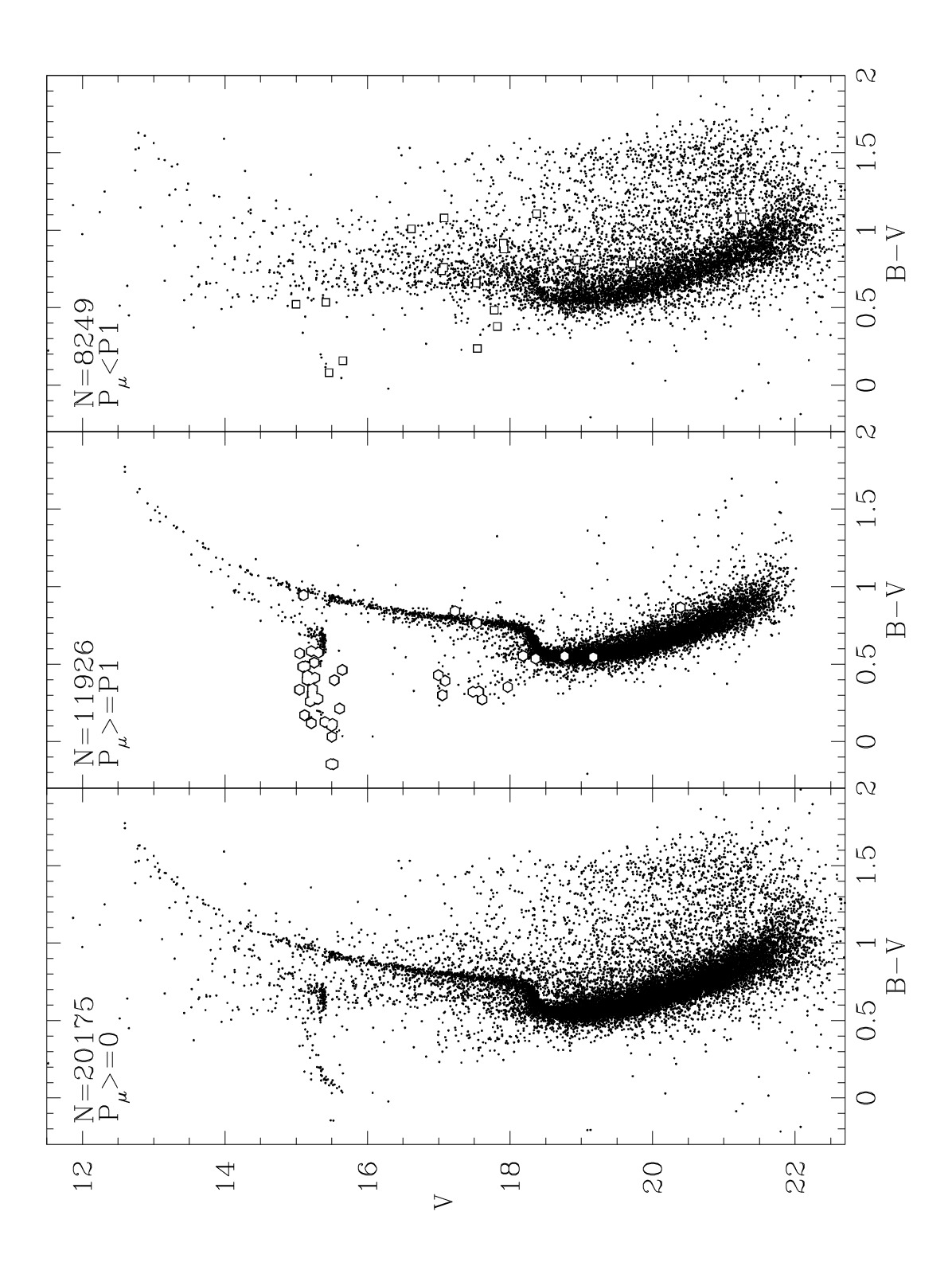}
    \caption{CMD of NGC~6362. All stars with PMs and colors measured (\emph{left}); stars with probabilities 
    $P_{\mu} \geq P1$ (\emph{middle}); stars with probabilities $P_{\mu} < P1$, where $P1=70\%$ 
    (\emph{right}).}
    \label{fig:ngc6362_cmd}
\end{figure*}

\begin{figure*}
	\includegraphics[width=0.55\textwidth, angle=-90]{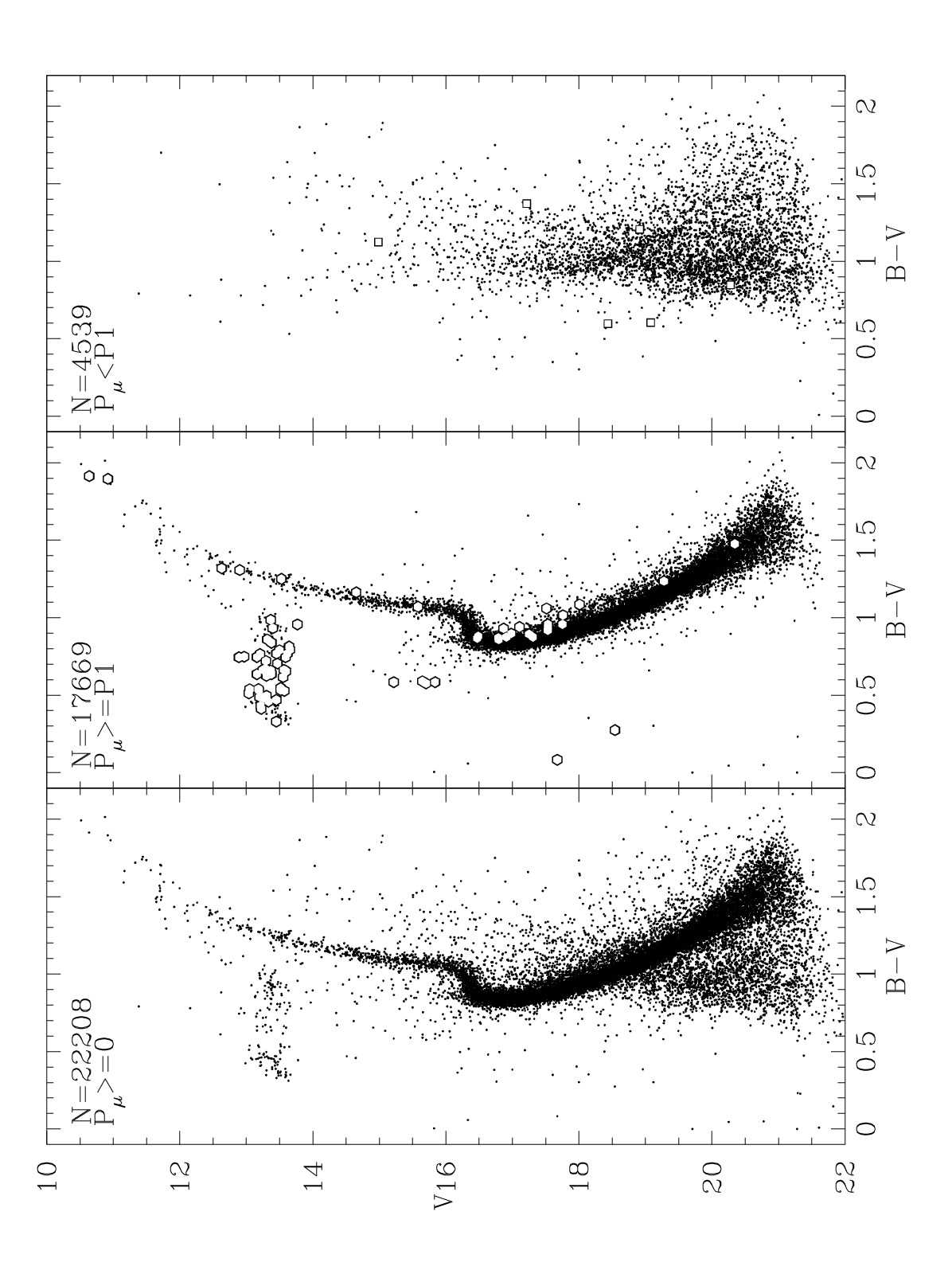}
    \caption{CMD of M4. All stars with PMs and colors measured (\emph{left}); stars with probabilities 
    $P_{\mu} \geq P1$ (\emph{middle}); stars with probabilities $P_{\mu} < P1$, where $P1=70\%$ 
    (\emph{right}).}
    \label{fig:m4_cmd}
\end{figure*}

\begin{figure*}
	\includegraphics[width=0.55\textwidth, angle=-90]{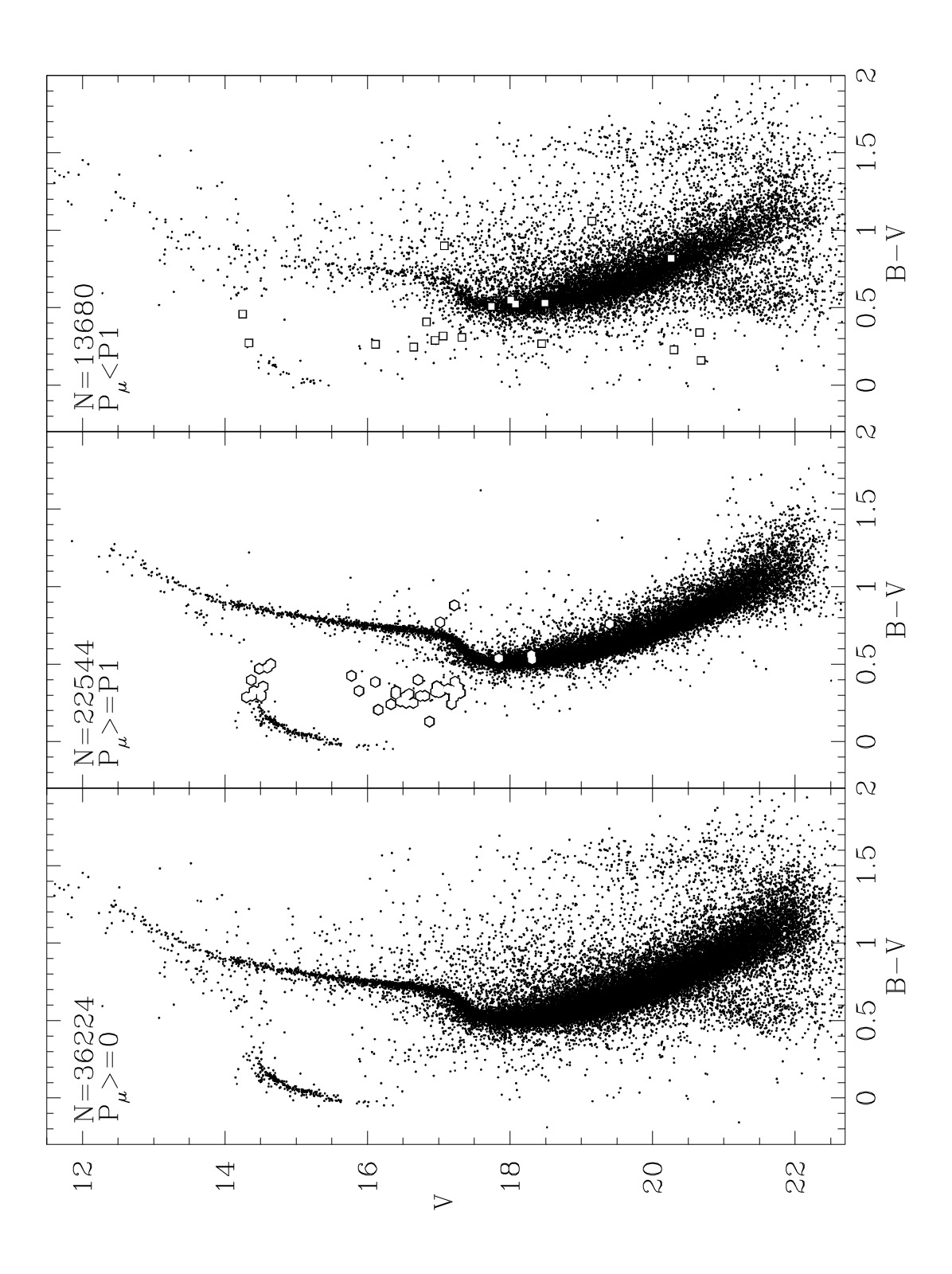}
    \caption{CMD of M55. All stars with PMs and colors measured (\emph{left}); stars with probabilities 
    $P_{\mu} \geq P1$ (\emph{middle}); stars with probabilities $P_{\mu} < P1$, where $P1=70\%$ 
    (\emph{right}).}
    \label{fig:m55_cmd}
\end{figure*}

\begin{figure*}
	\includegraphics[width=0.55\textwidth, angle=-90]{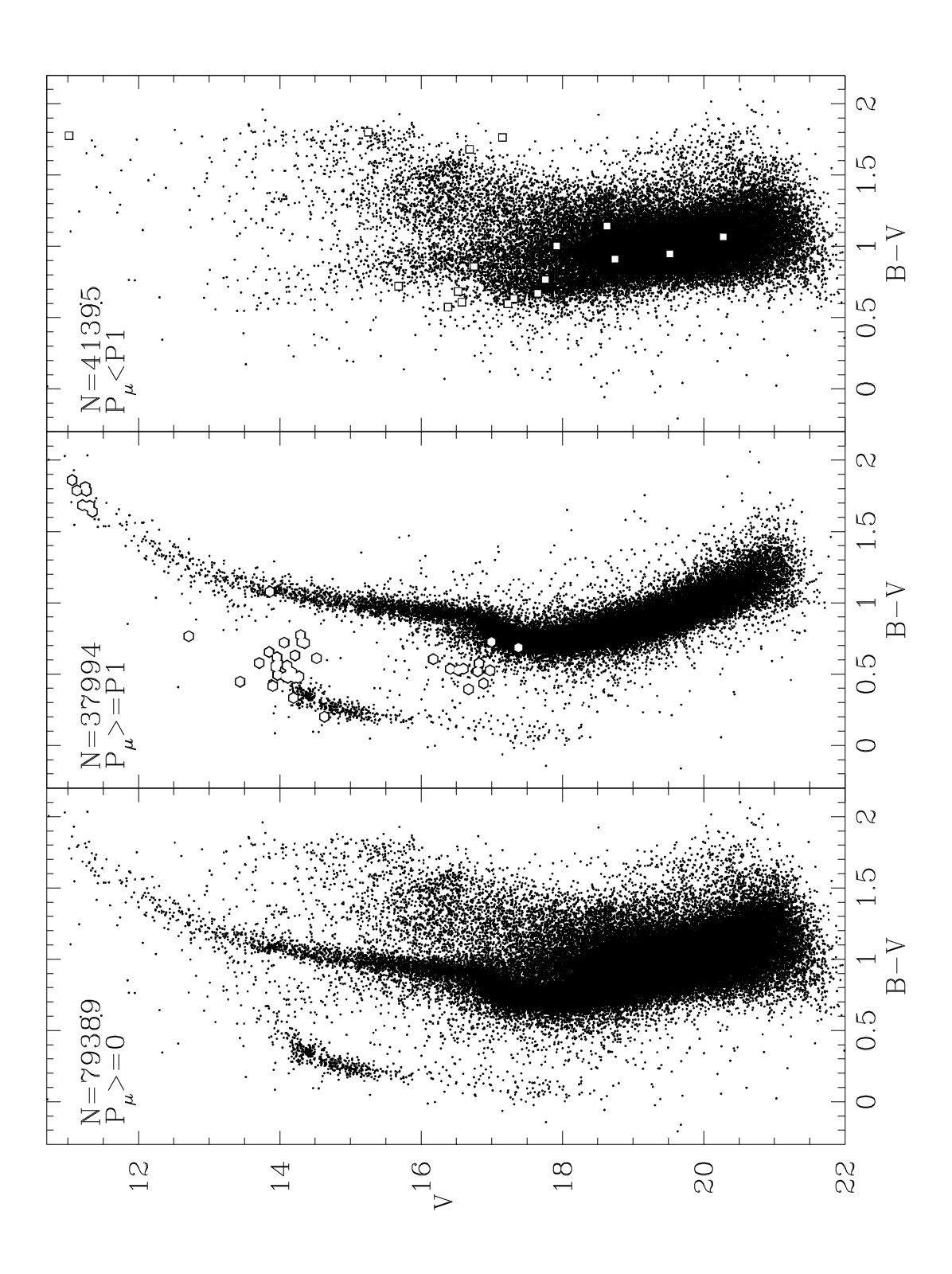}
    \caption{CMD of M22. All stars with PMs and colors measured (\emph{left}); stars with probabilities 
    $P_{\mu} \geq P1$ (\emph{middle}); stars with probabilities $P_{\mu} < P1$, where $P1=70\%$ 
    (\emph{right}).}
    \label{fig:m22_cmd}
\end{figure*}

\begin{figure*}
	\includegraphics[width=0.55\textwidth, angle=-90]{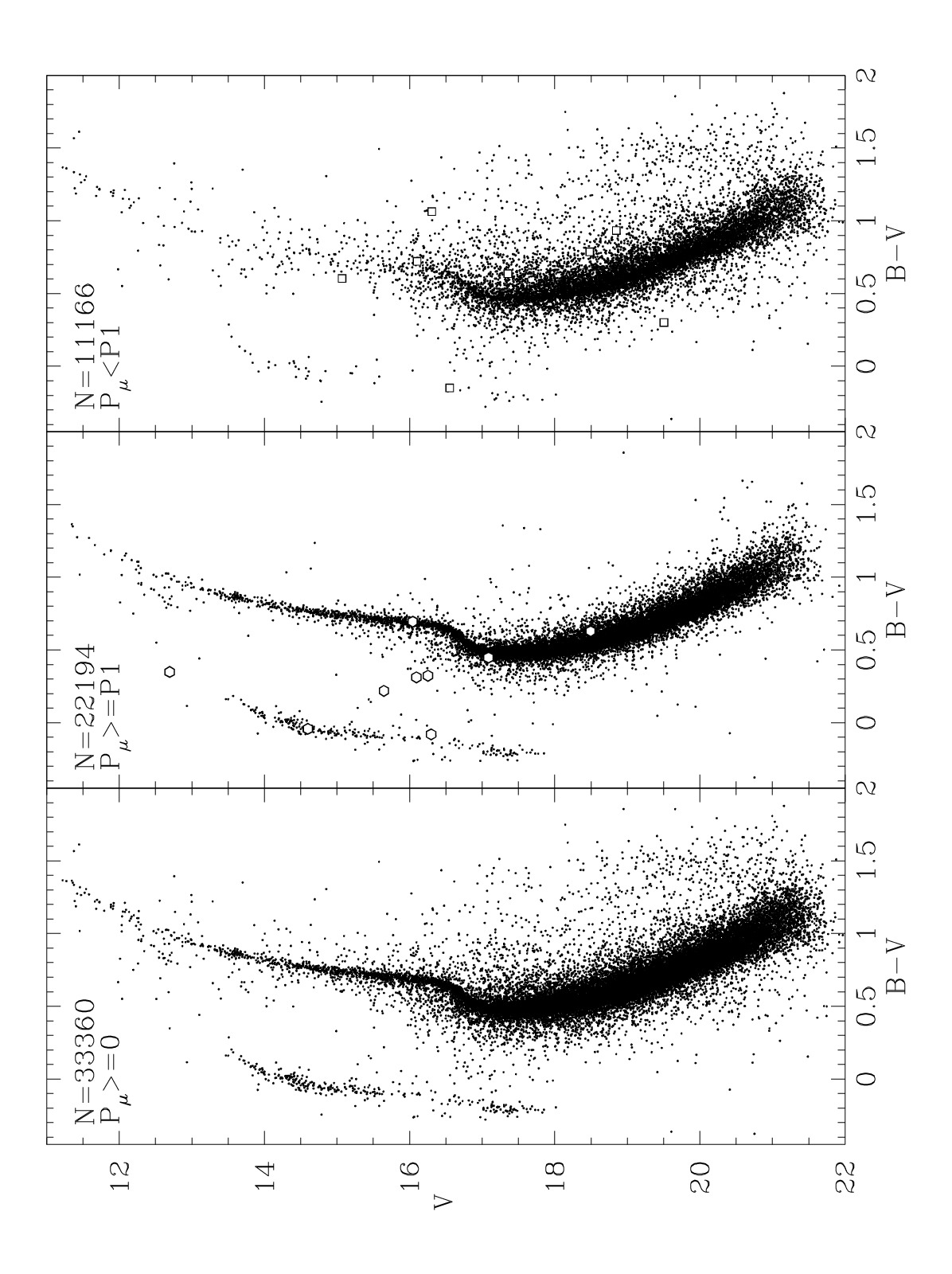}
    \caption{CMD of NGC~6752. All stars with PMs and colors measured (\emph{left}); stars with probabilities 
    $P_{\mu} \geq P1$ (\emph{middle}); stars with probabilities $P_{\mu} < P1$, where $P1=70\%$ 
    (\emph{right}).}
    \label{fig:ngc6752_cmd}
\end{figure*}

\begin{figure*}
	\includegraphics[width=0.55\textwidth, angle=-90]{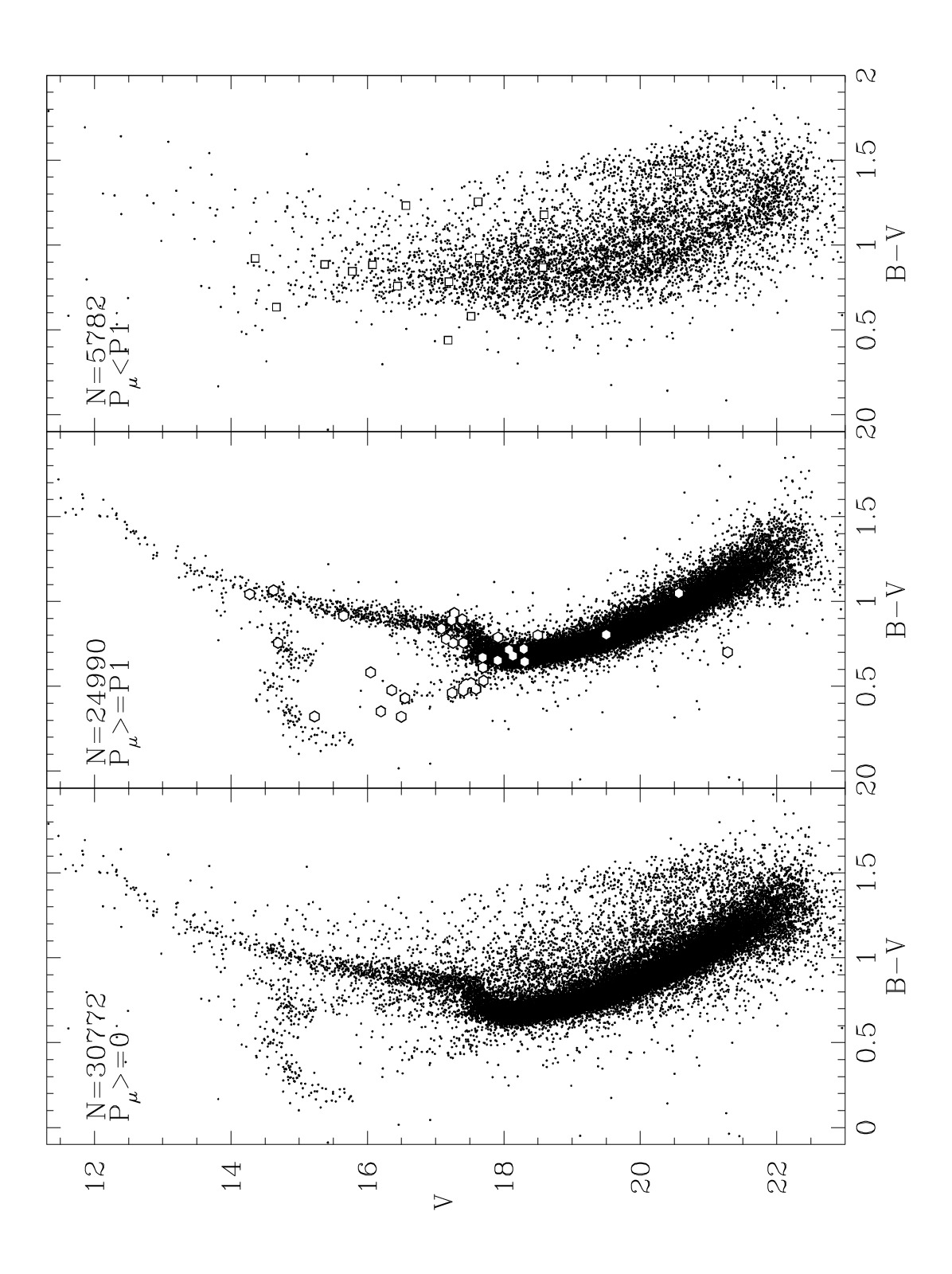}
    \caption{CMD of NGC~3201. All stars with PMs and colors measured (\emph{left}); stars with probabilities 
    $P_{\mu} \geq P1$ (\emph{middle}); stars with probabilities $P_{\mu} < P1$, where $P1=70\%$ 
    (\emph{right}).}
    \label{fig:ngc3201_cmd}
\end{figure*}

\begin{figure*}
	\includegraphics[width=0.55\textwidth, angle=-90]{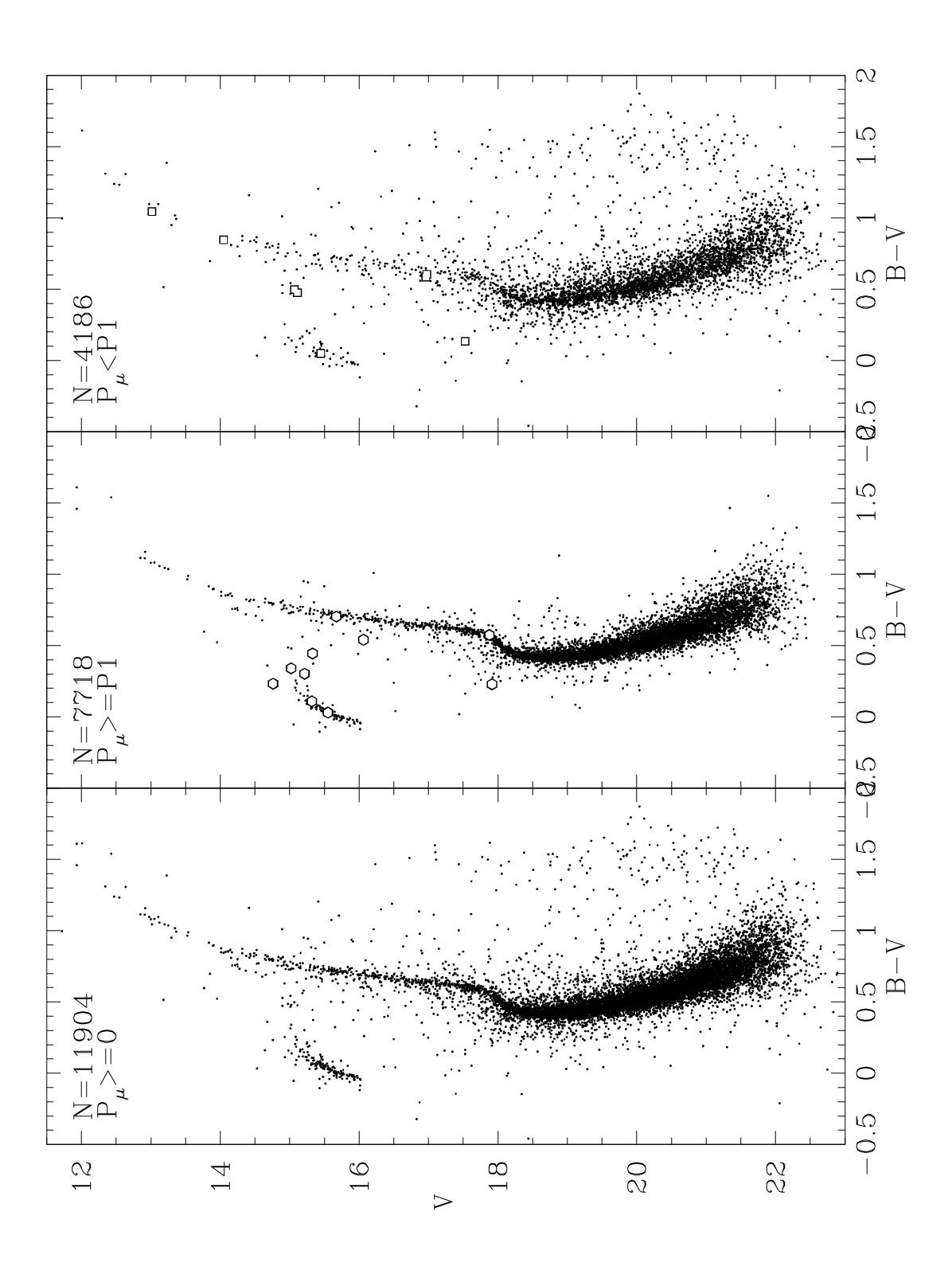}
    \caption{CMD of M30. All stars with PMs and colors measured (\emph{left}); stars with probabilities 
    $P_{\mu} \geq P1$ (\emph{middle}); stars with probabilities $P_{\mu} < P1$, where $P1=70\%$ 
    (\emph{right}).}
    \label{fig:m30_cmd}
\end{figure*}

\begin{figure*}
	\includegraphics[width=0.55\textwidth, angle=-90]{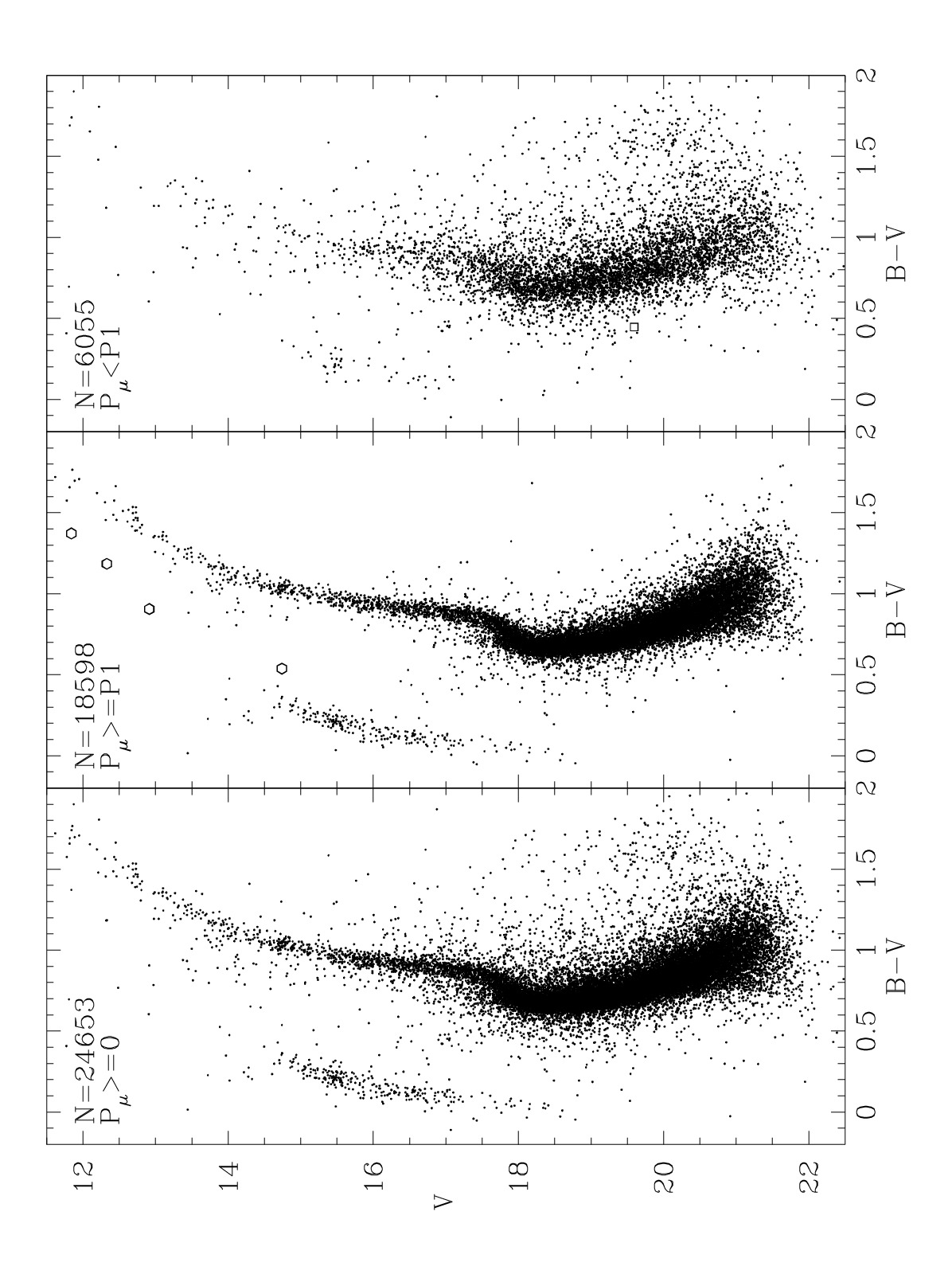}
    \caption{CMD of M10. All stars with PMs and colors measured (\emph{left}); stars with probabilities 
    $P_{\mu} \geq P1$ (\emph{middle}); stars with probabilities $P_{\mu} < P1$, where $P1=70\%$ 
    (\emph{right}).}
    \label{fig:m10_cmd}
\end{figure*}

\begin{figure*}
	\includegraphics[width=0.55\textwidth, angle=-90]{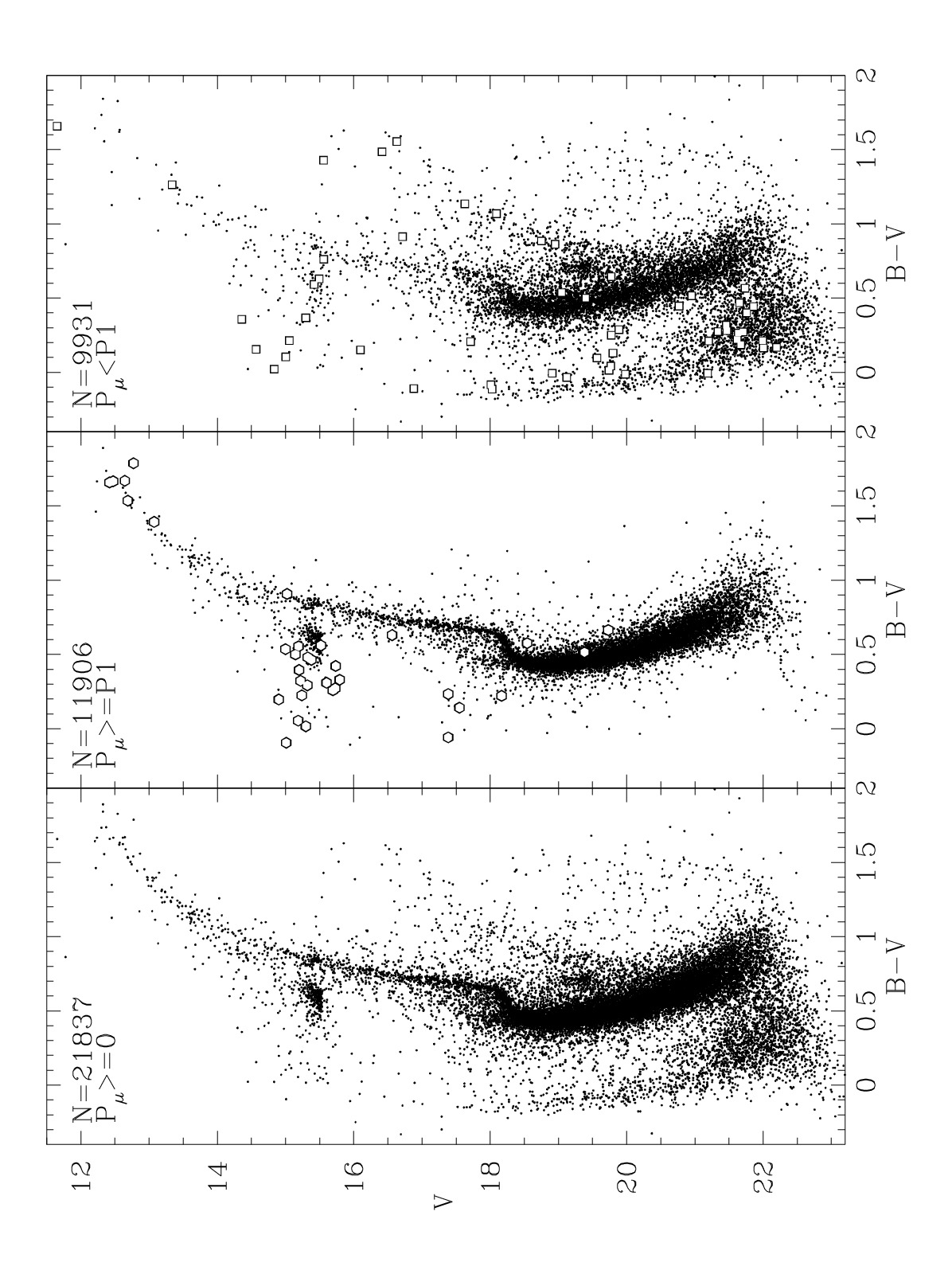}
    \caption{CMD of NGC~362. All stars with PMs and colors measured (\emph{left}); stars with probabilities 
    $P_{\mu} \geq P1$ (\emph{middle}); stars with probabilities $P_{\mu} < P1$, where $P1=70\%$ 
    (\emph{right}).}
    \label{fig:ngc362_cmd}
\end{figure*}

\begin{figure*}
	\includegraphics[width=0.55\textwidth, angle=-90]{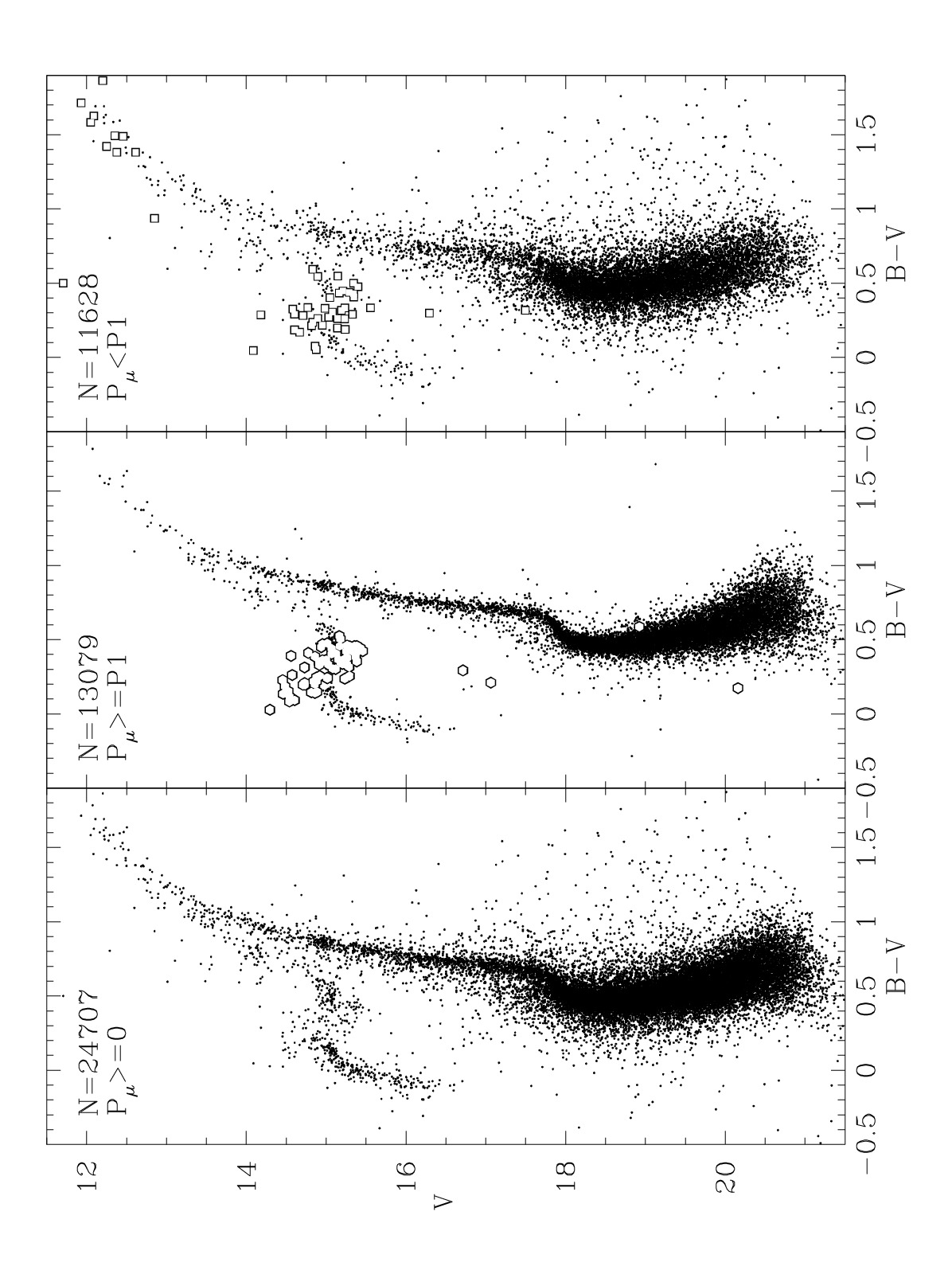}
    \caption{CMD of M5. All stars with PMs and colors measured (\emph{left}); stars with probabilities 
    $P_{\mu} \geq P1$ (\emph{middle}); stars with probabilities $P_{\mu} < P1$, where $P1=70\%$ 
    (\emph{right}).}
    \label{fig:m5_cmd}
\end{figure*}

\begin{figure*}
	\includegraphics[width=0.55\textwidth, angle=-90]{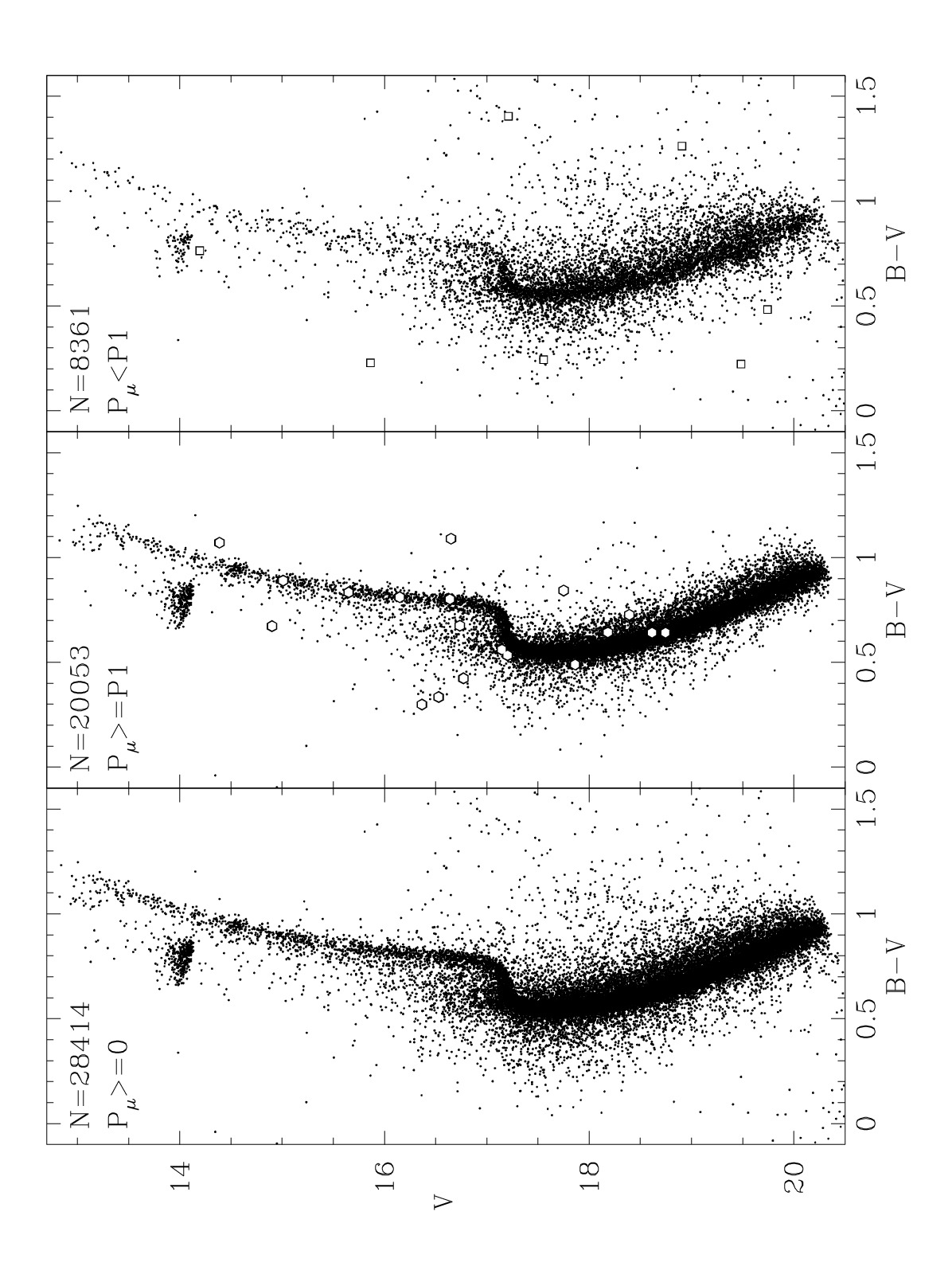}
    \caption{CMD of 47~Tuc field E. All stars with PMs and colors measured (\emph{left}); stars with 
    probabilities $P_{\mu} \geq P1$ (\emph{middle}); stars with probabilities $P_{\mu} < P1$, 
    where $P1=70\%$ (\emph{right}).}
    \label{fig:47tucE_cmd}
\end{figure*}

\begin{figure*}
	\includegraphics[width=0.55\textwidth, angle=-90]{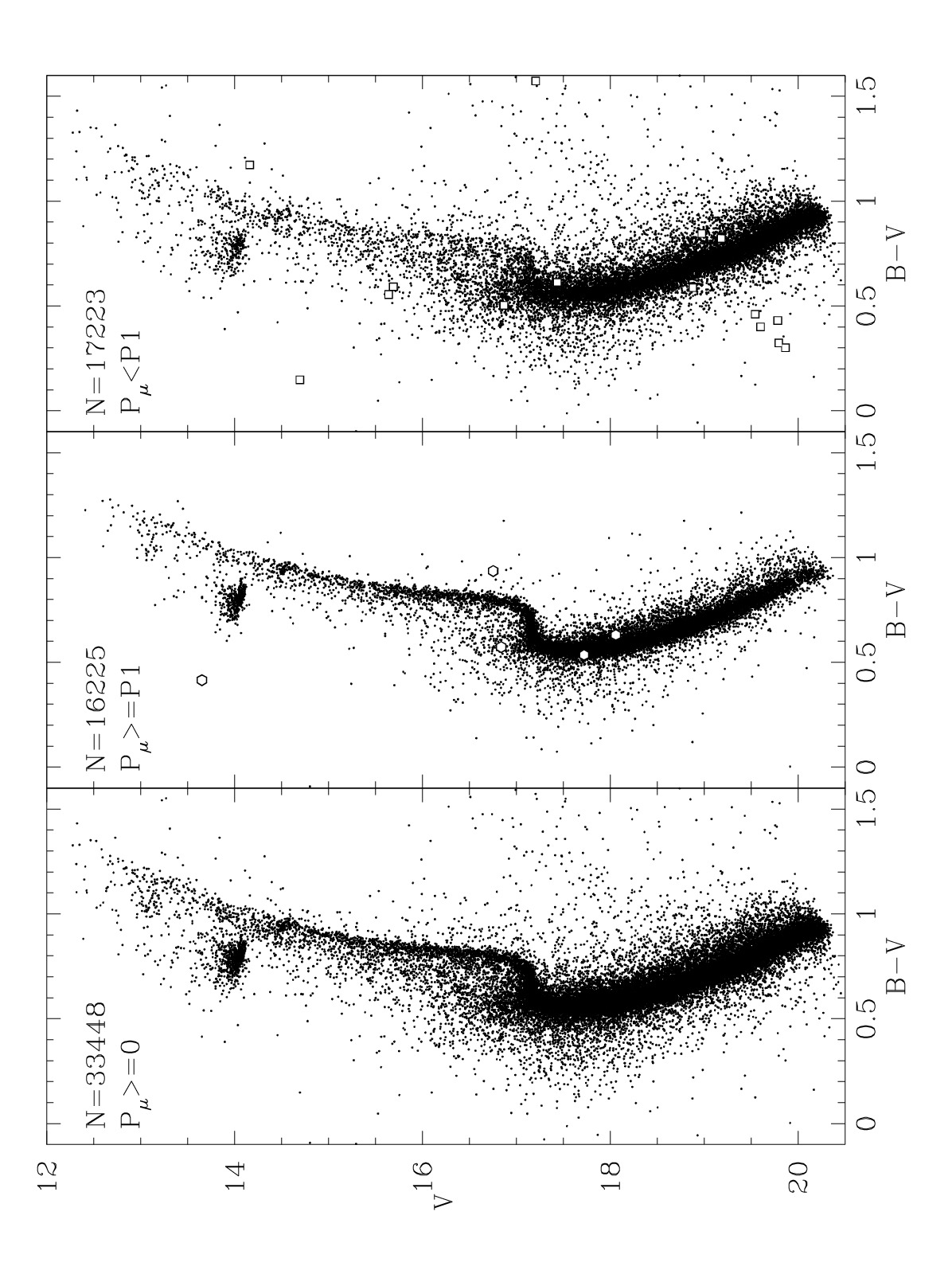}
    \caption{CMD of 47~Tuc field W. All stars with PMs and colors measured (\emph{left}); stars with 
    probabilities $P_{\mu} \geq P1$ (\emph{middle}); stars with probabilities $P_{\mu} < P1$, 
    where $P1=70\%$ (\emph{right}).}
    \label{fig:47tucW_cmd}
\end{figure*}

Fig.~\ref{fig:m12_cmd}--\ref{fig:47tucW_cmd} present the  CMDs of the analyzed cluster fields, 
and illustrate the effects of CMD cleaning based on membership probabilities. Since the membership 
status--based cleaning yields very similar results, the corresponding figures are not shown. Left 
panels include all stars with measured PM and ($B-V$) color. Middle panels include stars with 
$P_{\mu} \geq P1$ and right panels with $P_{\mu} < P1$, where $P1=70\%$. Additionally, in all middle 
and right panels we marked the positions of known variable stars for which we calculated proper 
motions. For most GCs, the list of variables was taken from the catalogue of \citet{clement2001} 
updated in $2010$\footnote{http://www.astro.utoronto.ca/$\textasciitilde$cclement/read.html}. The 
list of variables for NGC~362 was taken from the recent work of \citet{rozyczka2016}, and for 47~Tuc 
from \citet{kaluzny2013a}.

Not surprisingly, the CMD--cleaning procedure works best for GCs with VPDs well separated cluster 
and field stars (see Fig.~\ref{fig:m4_cmd}, \ref{fig:m22_cmd} and \ref{fig:ngc3201_cmd}). In the 
case of poorly separated GCs some field stars are remaining in the middle panels, and some likely 
cluster members (in particular, some MS stars) are counted as field objects and assigned to the 
right panels (see Fig.~\ref{fig:m12_cmd}, \ref{fig:m55_cmd} or \ref{fig:ngc6752_cmd}). For a~given 
cluster, the number of erroneous assignments might be reduced by suitable choice of P1, but in general 
such effects are inevitable while dealing with overlapping populations. The poorest results of CMD 
cleaning are observed in M5 and 47~Tuc~W, where PM uncertainties are particularly large (see 
Fig.~\ref{fig:m5_cmd} and \ref{fig:47tucW_cmd}). Moreover, in M5 many RR~Lyrae pulsators located 
in the HB region were assigned to field stars, although most likely they are cluster members 
measured with large uncertainties because of their location close to the cluster center. For 47~Tuc~W 
more stars were assigned to the field than to the cluster. The variables at the tip of the RGB are 
overexposed on many images, which again affects their uncertainties. Despite all these problems, 
the results presented in this Section prove that PMs are quite an efficient means to separate cluster 
members from field stars.

Membership status--based and membership probability--based cleanings yields similar results. For well 
separated clusters these are essentially identical but for poorly separated clusters membership 
probability--based cleaning rejects more stars fainter than $V \approx$19~mag than the other method. 
The reason for this was already pointed out in the previous section. Radius $r_{\mu}$ grows faster 
in the first method than in the second. The limit radius in both methods was not chosen accidentally. 
Assigning the star with PM less than three times its standard deviation to $mem=2$ class is very 
common limit in statistics and in this case should include $99.7\%$ of such stars. The value $P1=67\%$ 
is calculated based on comparison with radial velocities. But both these limits can be changed freely. 
Additionally, $P1$ can be adopted for each GC individually or instead of a~single value, it might be 
given as a~function, which would describe the $P_{\mu}(V)$ relation best. We have adopted $P1=const$. 
Membership probability--based cleaning is very sensitive to the number of stars in a~given magnitude 
interval and also to their spatial distribution on VPD (especially the field stars). That does not 
concern membership status--based cleaning were the average PMs are calculated for equal number of stars 
in the vincinity of the point $(0,0)$. Despite this, the first of these methods is more stable, because 
it is based on clear criteria and allows the smooth transition between magnitude intervals. In the 
second method, the intervals are not overlapping, and this is why the averages and their standard deviations 
might change significantly while changing the interval limits. Summarizing the above considerations, 
both cleaning methods have advantages and disadvantagesm, and neither is unquestionably better than the 
other. 

\section{Absolute proper motions}\label{sec:abspm}

\begin{figure}
	\includegraphics[width=1.2\columnwidth, angle=0]{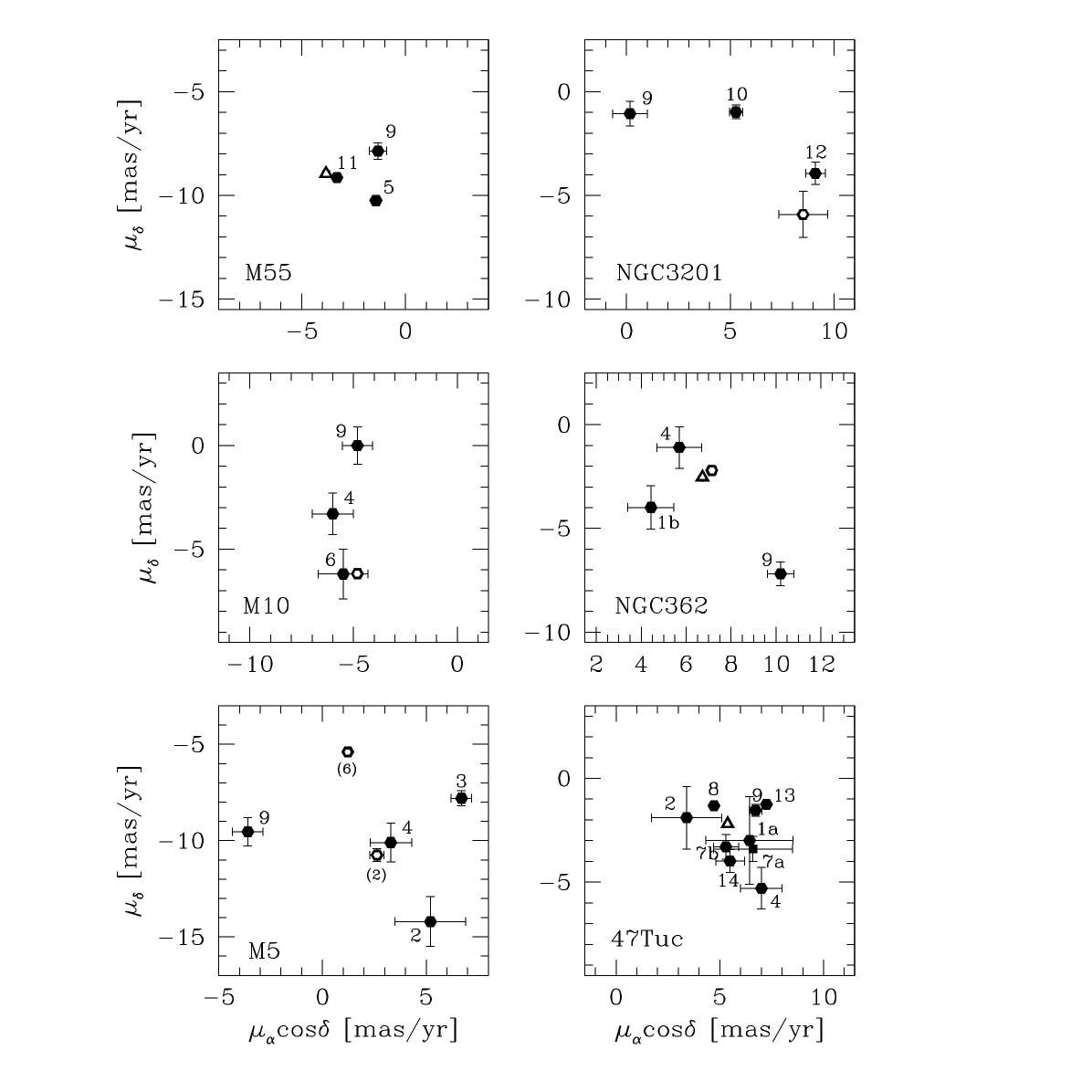}
    \caption{Absolute PMs for six GCs. Black and white triangles mark absolute PMs from this work calculated 
    relative to nearby galaxy and then corrected for its absolute PM from the literature. Black and white 
    points mark absolute PMs calculated basing on quasars identified in the field of a~given GC, where 
    $(6),(3)$~-- absolute PM calculated basing on six and three quasars, respectively. Black points mark 
    absolute PMs from the literature, where $1a,b$~-- \citet{tucholke1992a,tucholke1992b}, $2$~-- 
    \citet{cudworth1993},  $3$~-- \citet{scholz1996}, $4$~-- \citet{odenkirchen1997}, $5$~-- 
    \citet{dinescu1999}, $6$~-- \citet{chen2000}, $7a,b$~-- \citet{freire2001,freire2003}, $8$~-- 
    \citet{anderson2003}, $9$~-- \citet{dambis2006}, $10$~-- \citet{casetti-dinescu2007}, $11$~-- 
    \citet{zloczewski2011}, $12$~-- \citet{zloczewski2012}, $13$~-- \citet{cioni2016}, $14$~-- 
    \citet{watkins2016}.}
    \label{fig:abspm}
\end{figure}

The derived relative PMs allowed us to estimate the absolute proper motions for six GCs from 
our sample using close background galaxies (for M55, NGC~362 and 47~Tuc) or distant quasars 
(for NGC~3201, M10 and M5). The results are presented in Fig.~\ref{fig:abspm} and 
Table~\ref{tab:abs} together with absolute PMs from literature. 


To determine the relative motion of 47~Tuc and NGC~362 to the SMC we applied the same method as 
previously used when calculating membership probabilities (see Section~\ref{sec:memprob}). In the 
overlapping magnitude ranges we fitted two--dimensional Gaussian functions (given with Eq.~\ref{eq:g1} 
and \ref{eq:g2}) separately to the bulk of the cluster and SMC stars. The aperture radius for the 
cluster ($r_{GC}$) and SMC ($r_F$) were chosen to be $3\sigma$ and $1\sigma$, respectively. Next, 
we calculated the positions of Gaussians centroids, with that of the cluster located at $(0,0)$, 
as expected. 

The PM of a~GC relative to a~background object is simply the inverse of the PM of the object 
resulting from the cluster's VPD. We found the PM of 47~Tuc relative to SMC to be 
$(\mu_{\alpha}\cos\delta, \mu_{\delta}) = (4.54 \pm 0.02, -1.09 \pm 0.01)$~mas/yr, where the 
uncertainties were derived from formal errors of centroid location. This value agrees well with 
recent estimates by \citet{poleski2012}. Next, we corrected this value by the absolute PM of the SMC 
based on HST data from \citet{kallivayalil2013}, equal to $(0.772 \pm 0.063, -1.117 \pm 0.061)$~mas/yr 
and \citet{piatek2008}, equal to $(0.754 \pm 0.061, -1.252 \pm 0.058)$~mas/yr, but also background 
measurements from \citet{costa2009}, equal to $(1.03 \pm 0.29, -1.09 \pm 0.18)$~mas/yr, \citet{costa2011}, 
equal to $(0.93 \pm 0.14, -1.25 \pm 0.11)$~mas/yr, \citet{vieira2010}, equal to 
$(0.98 \pm 0.30, -1.01 \pm 0.29)$~mas/yr, \citet{cioni2016}, equal to $(1.16 \pm 0.07, -0.81 
\pm 0.07)$~mas/yr, and recent Tycho--Gaia Astrometric Solution (TGAS) PM from \citet{vandermarel2016}, 
equal to $(0.874 \pm 0.066, -1.229 \pm 0.047)$~mas/yr. Accounting for the mean absolute PM of the SMC 
based on above values from literature, we obtained a~mean absolute PM of 47~Tuc of $(5.376 \pm 0.032, 
-2.216 \pm 0.028)$~mas/yr. This result broadly agrees with earlier measurements.


An analogous calculation for NGC~362 gives the relative PM to SMC of $(5.76 \pm 0.01, -1.45 \pm 
0.01)$~mas/yr, and a~mean absolute PM of $(6.726 \pm 0.032, -2.563 \pm 0.028)$~mas/yr. Additionally, 
in the field of NGC~362 we found that the object \#$2302977$ is the quasar candidate J010239.8-705803 
from the largest existing quasar catalogue~-- The Million Quasars 
catalog\footnote{Available on: http://quasars.org/milliquas.htm} \citep[Milliquas,][]{flesch2015}. 
The redshift of this object is unknown but it has an assigned probability of being a~quasar of $96\%$. 
Based on the PM of this quasar the absolute PM of the cluster is $(7.15 \pm 0.19, -2.21 \pm 0.21)$~mas/yr. 
These values are equal within the errors, and they coincide with previous estimations, e.g. 
\citet{odenkirchen1997}. 

Since the bulk of the Sagittarius~dSph (Sgr-dSph) galaxy is not well pronounced in the VPD of M55 
as the SMC in the above two cases, we decided to use a~slightly different approach. We selected stars 
with $16<V<23$~mag lying within $1$~mas/yr from the mean motion of the bulk. Stars with PM errors larger 
than $10$~mas/yr were rejected from the sample. In the end, we obtain a sample of $309$ stars for which 
we measured mean motions. The resulting PM of M55 relative to Sgr~dSph was estimated to be $(-0.10 \pm 0.01, 
-7.44 \pm 0.01$~mas/yr. Following \citet{sohn2015}, we adopted the average PM of the mass center of Sgr~dSph 
to be $(-2.82 \pm 0.11, -1.51 \pm 0.14)$~mas/yr, and the resulting  absolute PM of M55 is 
$(-3.82 \pm 0.11, -8.95 \pm 0.14)$~mas/yr. The latter authors indicate that, because of perspective 
effects, measuring PMs in different parts of the extremely stretched Sgr~dSph gives different results. 
This might be the reason why our value differs from $(-3.31 \pm 0.10, -9.14 \pm 0.15)$~mas/yr obtained 
by \citet{zloczewski2011} for the absolute PM of M55.

We identified objects \#$260440$ and \#$260472$ in the field of M10 to be the  quasars 
SDSS~J165714.34-041625.9 and SDSS~J165713.57-041620.7 from the Milliquas catalogue. Their redshifts 
are $z \approx$1.4 and $2.5$, respectively. In the catalogue they have assigned probability of being 
a~quasar equal to $95\%$ and $96\%$. The weighted mean motion of these objects results in an absolute 
PM of M10 of $(-4.82 \pm 0.15, -6.18 \pm 0.13)$~mas/yr. This value coincides within the errors with 
the estimate given by \citet{chen2000}. 

In the field of M5 we found six Milliquas objects. We identified object \#$2600420$ as J151840.4+015352, 
\#$4500589$ as J151819.4+015918, \#$4401792$ as J151819.1+020152, \#$4201481$ as SDSS~J151810.74+021257.2, 
\#$1600139$ as SDSS~J151857.62+015345.9 and \#$3100796$ as SDSS~J151824.02+021347.6. For the first three 
objects there is no redshift given in the catalogue, and the assigned probabilities of being a~quasar 
are $83\%$, $80\%$ and $86\%$, respectively. Redshifts of the remaining objects are equal to $2.8$, $1.0$ 
and $1.2$, respectively, and the assigned probabilities range from $98\%$ to $99\%$. The absolute PM of 
M5 based on the weighted mean of motions of all six quasars is $(1.22 \pm 0.09, -5,40 \pm 0.08)$~mas/yr. 
We found that the object SDSS~J151810.74+021257.2 in the USNO catalogue is identified with object 
0922-0339533 which has significant absolute PM, so most probably it is a misidentification. The absolute 
PM of M5 based only on the two quasars for which the redshifts are known is $(2.63 \pm 0.34, -10.75 \pm 
0.34)$~mas/yr. The discrepancy between both values comes from the fact that only two quasars with known 
redshifts have significant PMs (\#$1600139$ and \#$3100796$), the remaining four have PMs close to zero. 


The object \#$337214$ in the field of NGC~3201 in \citet{kaluzny2016} was found to be a~quasar with 
a~redshift $z \approx$0.5 and is an X--ray counterpart \citep[][, object J101715.62-462253.2]{motch2016}. 
It does not appear in the Milliquas catalogue. Based on its relative PM, the absolute PM of NGC~3201 
is $(8.52 \pm 1.17, -5.92 \pm 1.11)$~mas/yr. The absolute PM of NGC~3201 was measured by 
\citet{zloczewski2012} with a result of $(9.11 \pm 0.47, -3.94 \pm 0.54)$~mas/yr. The difference 
between these two result is statistically insignificant, but both of these estimates are significantly 
different from values presented by \citet{casetti-dinescu2007} or \citet{dambis2006}.

In most cases, our absolute PMs agree within the errors with the previous estimations. Quasars provide 
the best and most straightforward means to determine the absolute PMs of GCs. Unfortunately, only very 
few of them have been found so far in cluster fields, especially in southern hemisphere, where optical 
quasar catalogues are far from being complete.

\begin{table*}[p]
	\centering
	\caption{Absolute PMs for six GCs.}
	\label{tab:abs}
	\begin{tabular}{lrrrrl} \hline
	 & \multicolumn{2}{c}{This work} & \multicolumn{2}{c}{Literature} & \\
	ID & $\mu_{\alpha}cos\delta$ [mas/yr] & $\mu_{\delta}$ [mas/yr] & $\mu_{\alpha}cos\delta$ [mas/yr] & 
	$\mu_{\delta}$ [mas/yr] & Reference \\ \hline 
	M55 & $-3.82 \pm 0.11$ & $-8.95 \pm 0.14$ & $-1.42 \pm 0.62$ & $-10.25 \pm 0.64$ & \citet{dinescu1999} \\
	    &                  &                  & $-1.33 \pm 0.42$ &  $-7.86 \pm 0.40$ & \citet{dambis2006} \\
	    &                  &                  & $-3.31 \pm 0.10$ &  $-9.14 \pm 0.15$ & \citet{zloczewski2011} \\
	NGC~3201 & $8.52 \pm 1.17$ & $-5.92 \pm 1.11$ & $0.17 \pm 0.83$ & $-1.06 \pm 0.60$ & \citet{dambis2006} \\
		 &                 &                  & $5.28 \pm 0.32$ & $-0.98 \pm 0.33$ & \citet{casetti-dinescu2007} \\
	M10 & $-4.82 \pm 0.15$ & $-6.18 \pm 0.13$ & $-6.00 \pm 1.00$ & $-3.30 \pm 1.00$ & \citet{odenkirchen1997} \\
	    &                  &                  & $-5.50 \pm 1.20$ & $-6.20 \pm 1.20$ & \citet{chen2000} \\
	    &                  &                  & $-4.81 \pm 0.72$ & $-0.01 \pm 0.90$ & \citet{dambis2006} \\
	NGC~362 & $6.726 \pm 0.032$  & $-2.563 \pm 0.028$ $^{1}$  & $4.43 \pm 1.02$ & $-3.99 \pm 1.04$ & \citet{tucholke1992b} \\
		& $7.15 \pm 0.19$ & $-2.21 \pm 0.21$ $^{2}$ & $5.70 \pm 1.00$ & $-1.10 \pm 1.00$ & \citet{odenkirchen1997} \\
		&                     &                      & $10.20 \pm 0.59$ & $-7.19 \pm 0.57$ & \citet{dambis2006} \\
	M5 & $2.63 \pm 0.34$ & $-10.75 \pm 0.34$ $^{3}$ &  $5.20 \pm 1.70$ & $-14.20 \pm 1.30$ & \citet{cudworth1993} \\
	   & $1.22 \pm 0.09$ & $-5.40 \pm 0.08$ $^{4}$ &  $6.70 \pm 0.50$ & $-7.80 \pm 0.40$ & \citet{scholz1996} \\
	   &                 &                  &  $3.30 \pm 1.00$ & $-10.10 \pm 1.00$ & \citet{odenkirchen1997} \\
	   &                 &                  & $-3.60 \pm 0.74$ & $-9.54 \pm 0.73$ & \citet{dambis2006} \\
	47~Tuc & $5.376 \pm 0.032$ & $-2.216 \pm 0.028$ & $6.43 \pm 2.10$ & $-2.99 \pm 2.11$ & \citet{tucholke1992a} \\
	       &                    &                     & $3.40 \pm 1.70$ & $-1.90 \pm 1.50$ & \citet{cudworth1993} \\
	       &                    &                     & $7.00 \pm 1.00$ & $-5.30 \pm 1.00$ & \citet{odenkirchen1997} \\
	       &                    &                     & $6.60 \pm 1.90$ & $-3.40 \pm 0.60$ & \citet{freire2001} \\
	       &                    &                     & $5.30 \pm 0.60$ & $-3.30 \pm 0.60$ & \citet{freire2003} \\
	       &                    &                     & $4.716 \pm 0.035$ & $-1.325 \pm 0.021$ & \citet{anderson2003}\\
	       &                    &                     & $6.73 \pm 0.30$ & $-1.54 \pm 0.29$ & \citet{dambis2006}\\
	       &                    &                     & $6.90 \pm 1.00$ & $-2.1 \pm 1.00$ & \citet{girard2011}$^{a}$ \\
	       &                    &                     & $4.41 \pm 0.67$ & $-1.12 \pm 0.55$ & \citet{poleski2012}$^{a}$ \\
	       &                    &                     & $7.26 \pm 0.03$ & $-1.25 \pm 0.03$ & \citet{cioni2016} \\ 
	       &                    &                     & $5.50 \pm 0.70$ & $-3.99 \pm 0.55$ & \citet{watkins2016} \\ \hline
	\end{tabular}\\
$^{1}$ PM relative to SMC and corrected for its absolute PM. \hspace{0.5cm}
$^{2}$ PM based on single quasar.\\
$^{3}$ PM based on two quasars. \hspace{0.5cm}
$^{4}$ PM based on six quasars.\\
$^{a}$ PM relative to SMC.
\end{table*}

\section{Summary}\label{sec:summary}

Based on data collected between $1997$ to $2015$ we obtained proper motions of over $446\,000$ stars 
in the fields of twelve nearby Galactic globular clusters: M12, NGC~6362, M4, M55, M22, NGC~6752, 
NGC~3201, M30, M10, NGC~362, M5, and 47~Tuc. The measurements were made using procedures similar to 
those used by \citet{anderson2006} and \citet{zloczewski2011, zloczewski2012}. Each of those stars was 
assigned to one of the three membership classes, and its membership probability was calculated. Both 
approaches allow to efficiently separate field stars from cluster members on CMDs. For six GCs the 
absolute PMs were obtained. The catalogues of the derived PMs will be freely accessible from 
\emph{http://case.camk.edu.pl/} and VizieR/CDS. 

\section*{Acknowledgements}
\label{sec:acn} 

WN, JK, MR and WP were partly supported by the grant DEC-2012/05/B/ST9/03931 from the Polish 
National Science Center.







\clearpage
\appendix

\section{Some extra material}
\label{sec:extra}

\begin{figure}
	\includegraphics[width=\columnwidth]{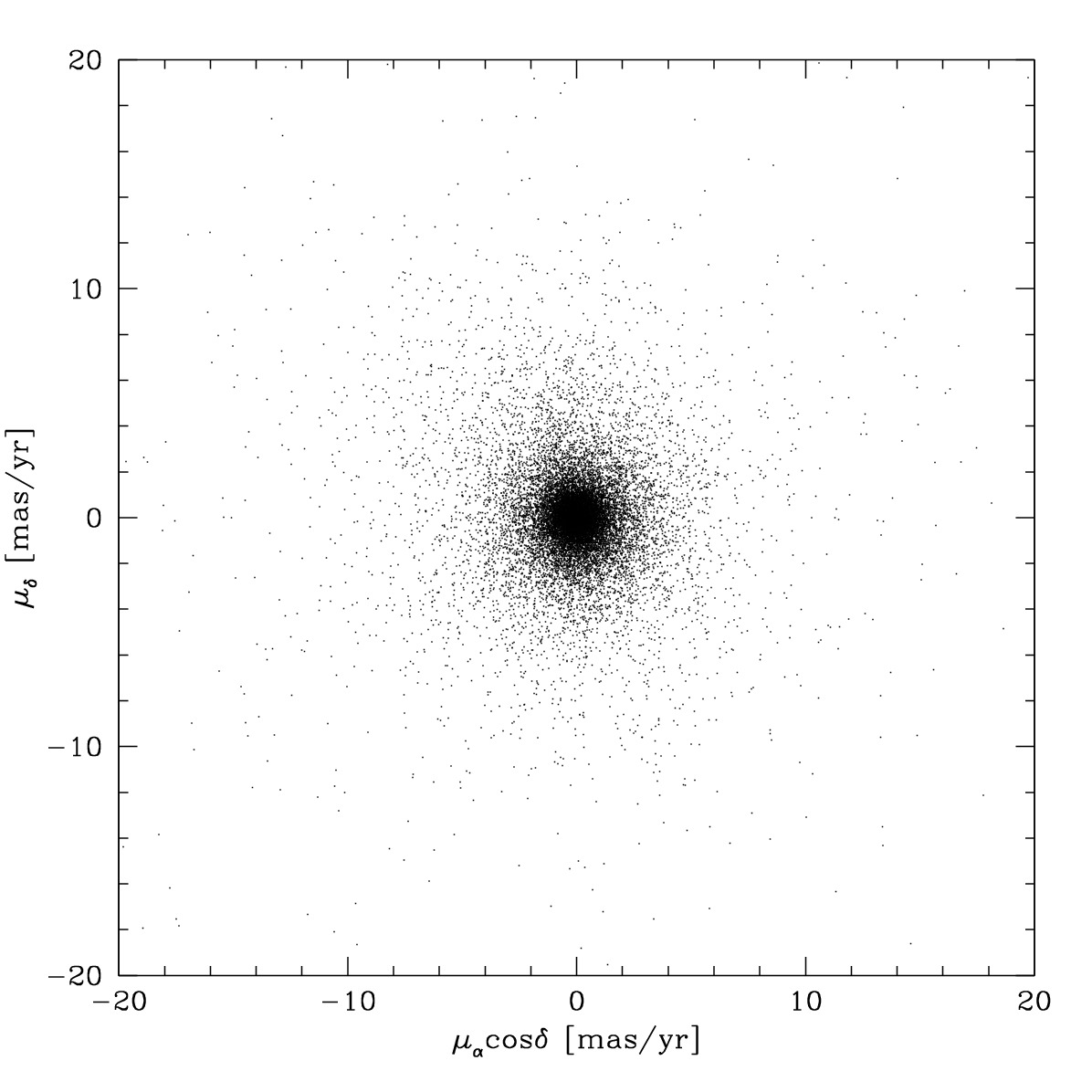}
    \caption{Vector point diagram (VPD) for M12.}
    \label{fig:m12vpd}
\end{figure}

\begin{figure}
	\includegraphics[width=\columnwidth]{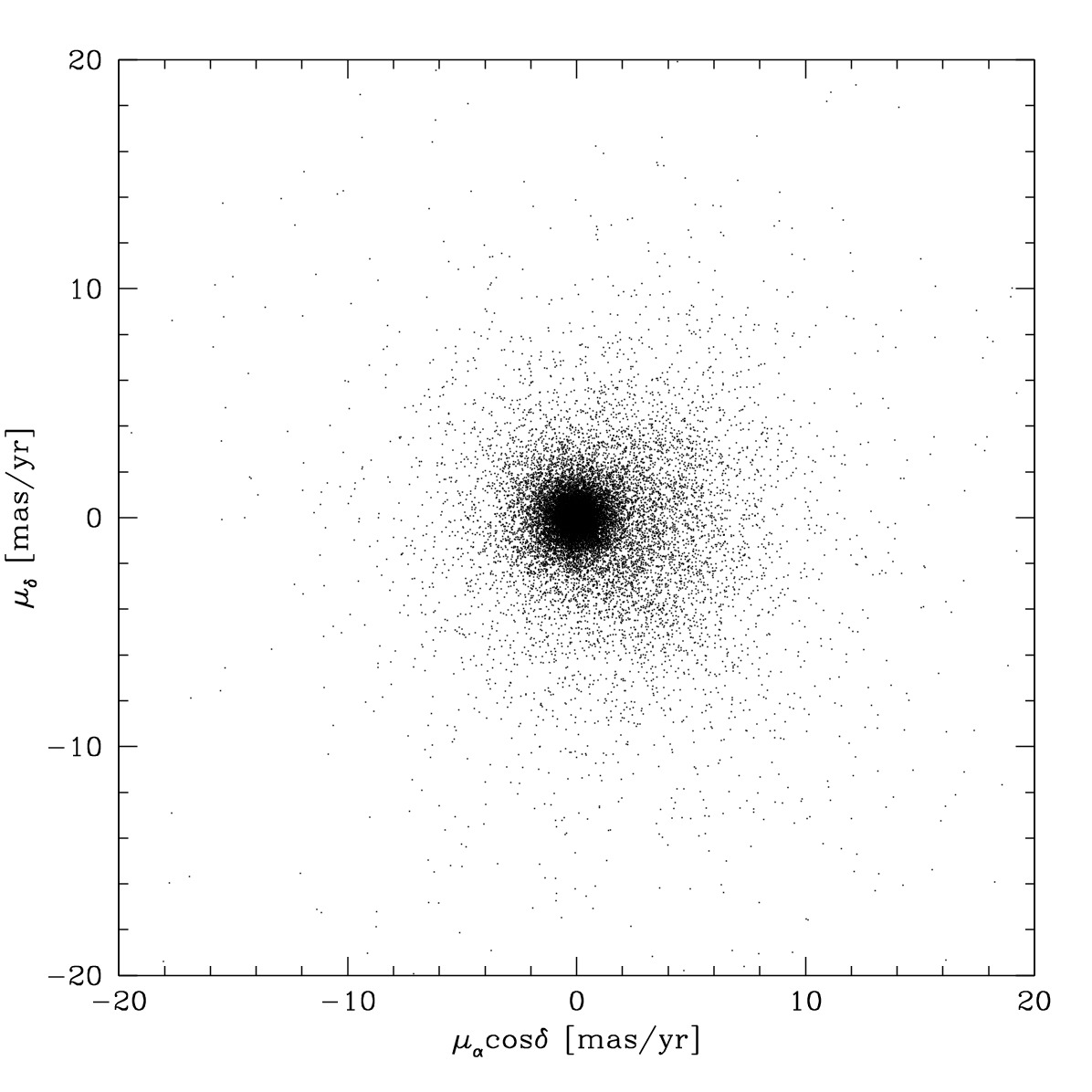}
    \caption{VPD for NGC~6362.}
    \label{fig:ngc6362vpd}
\end{figure}

\begin{figure}
	\includegraphics[width=\columnwidth]{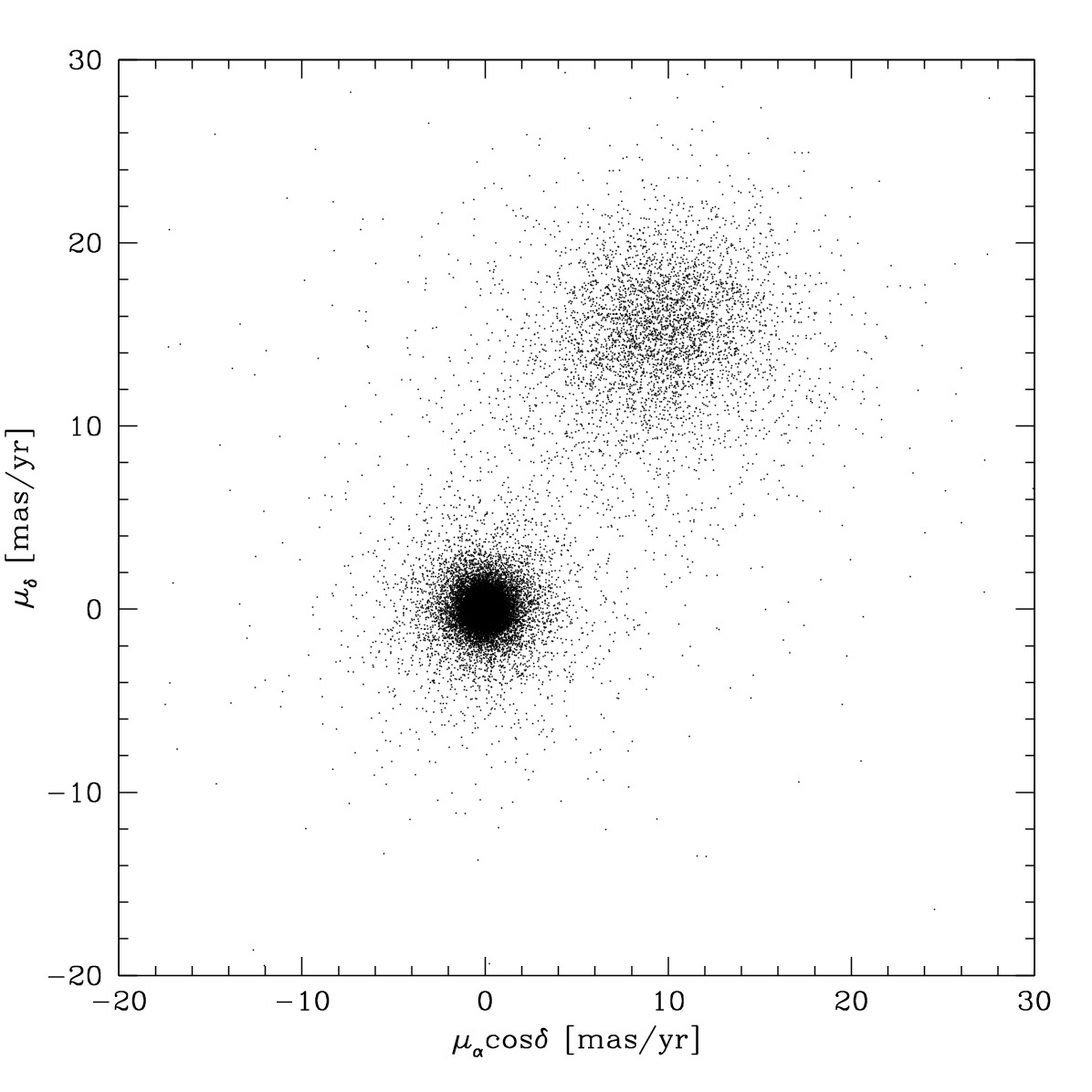}
    \caption{VPD for M4.}
    \label{fig:m4vpd}
\end{figure}

\begin{figure}
	\includegraphics[width=\columnwidth]{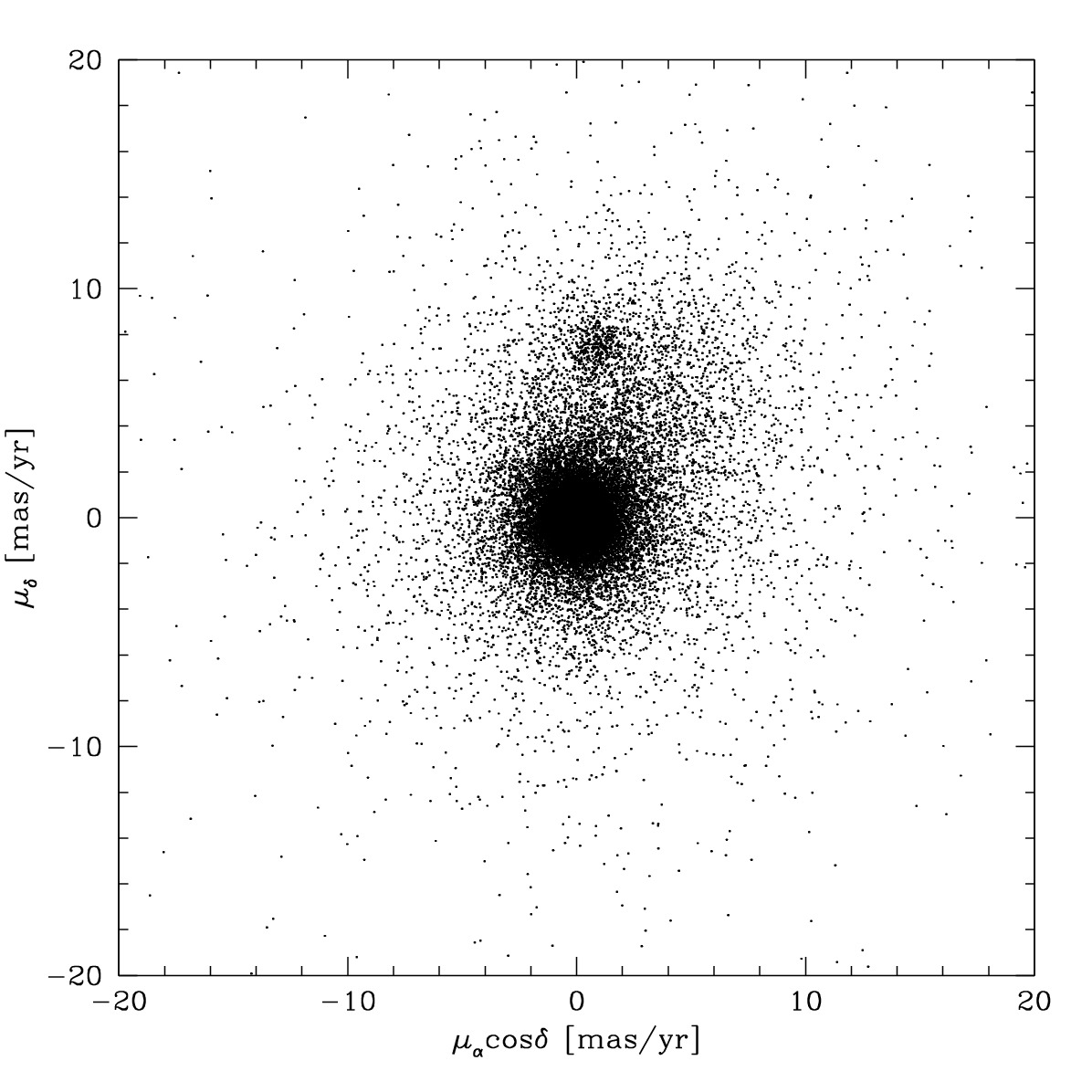}
    \caption{VPD for M55.}
    \label{fig:m55vpd}
\end{figure}

\begin{figure}
	\includegraphics[width=\columnwidth]{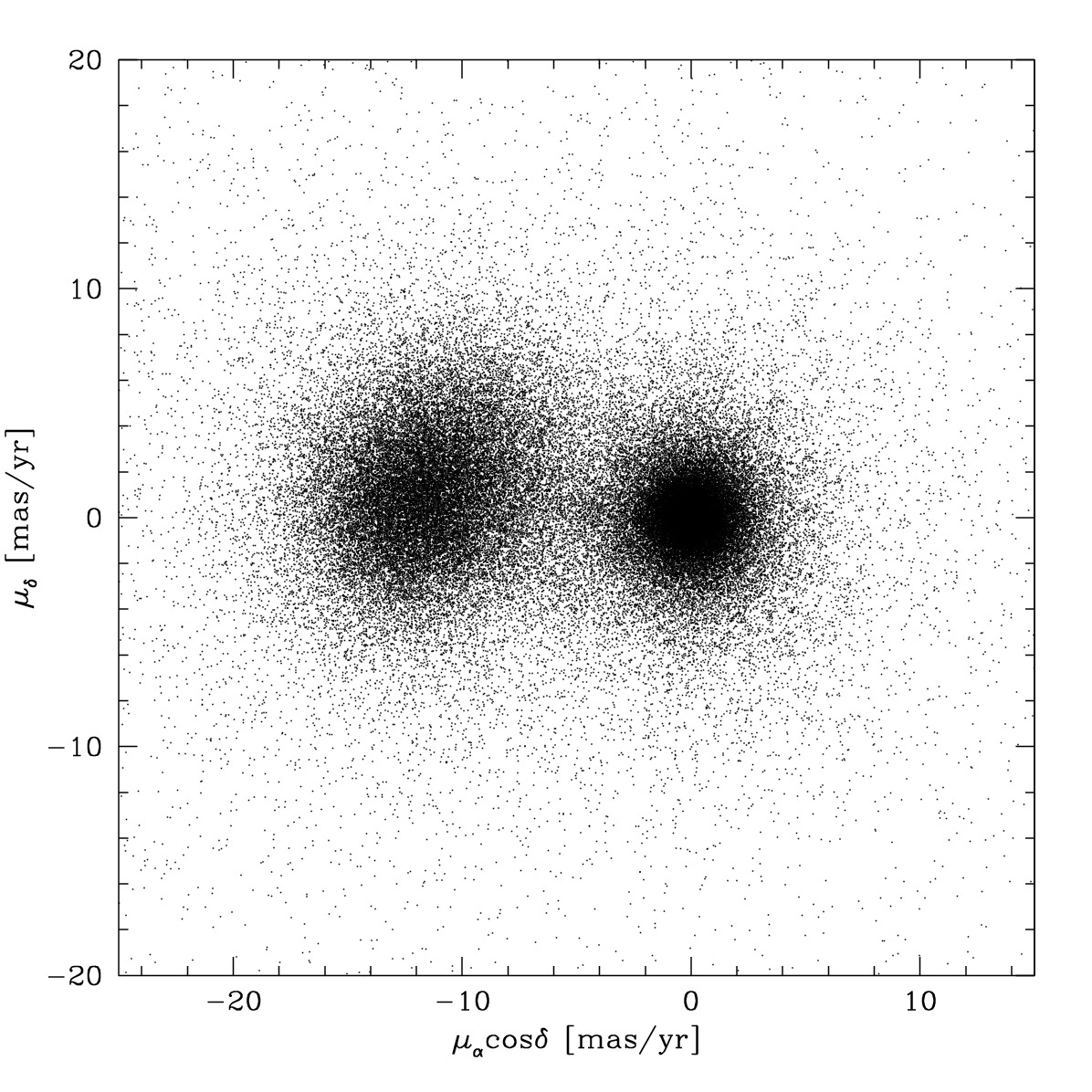}
    \caption{VPD for M22.}
    \label{fig:m22vpd}
\end{figure}

\begin{figure}
	\includegraphics[width=\columnwidth]{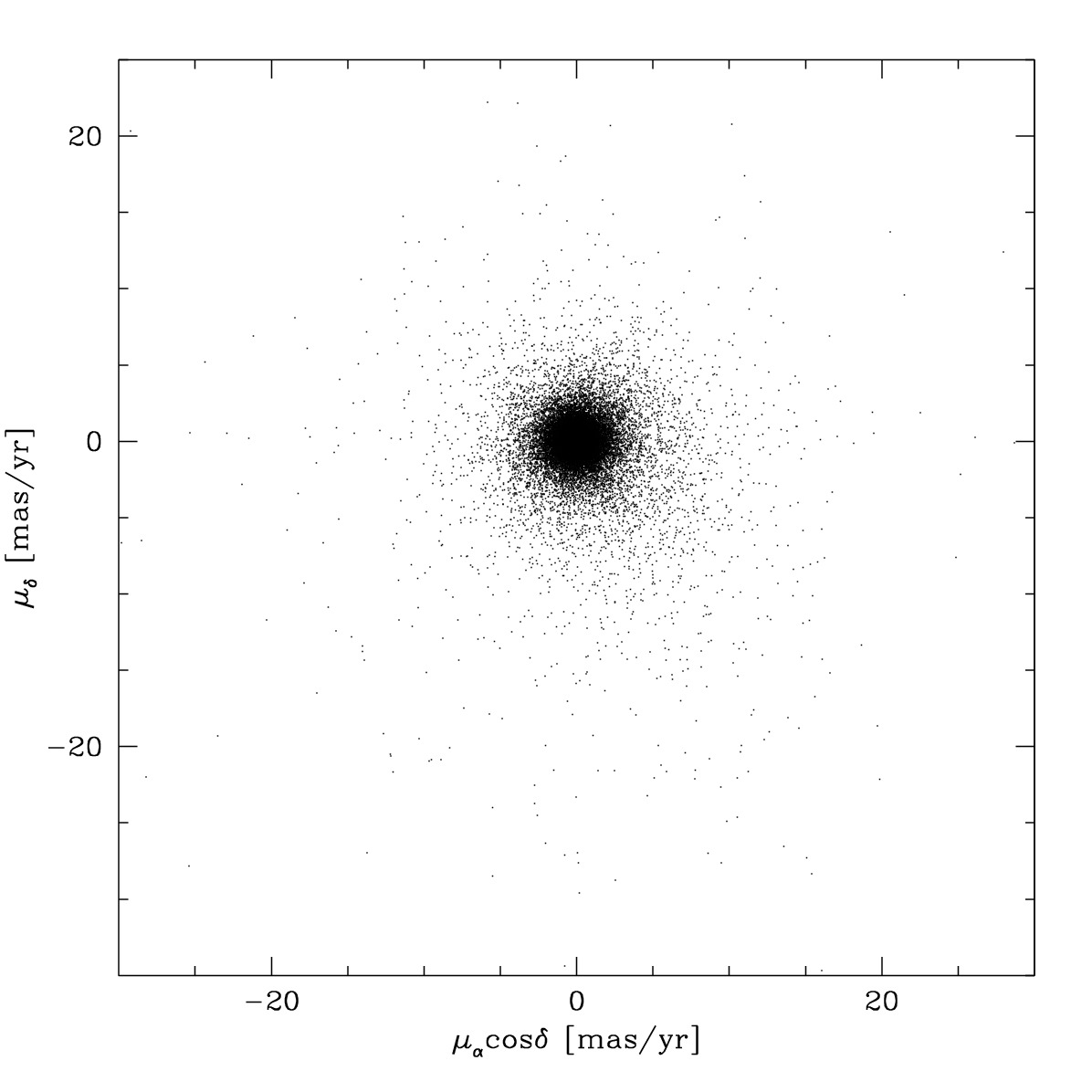}
    \caption{VPD for NGC~6752.}
    \label{fig:ngc6752vpd}
\end{figure}

\begin{figure}
	\includegraphics[width=\columnwidth]{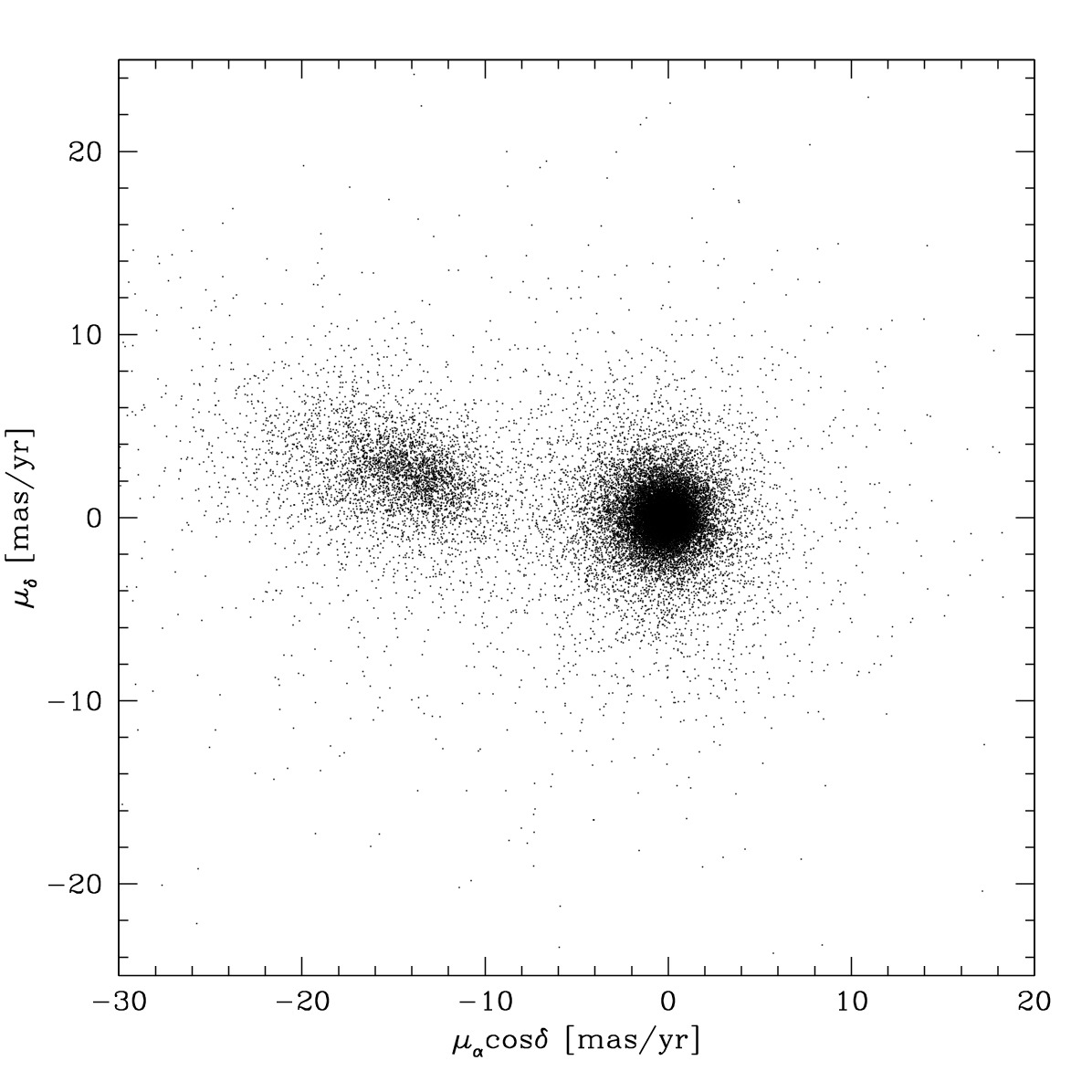}
    \caption{VPD for NGC~3201.}
    \label{fig:ngc3201vpd}
\end{figure}

\begin{figure}
	\includegraphics[width=\columnwidth]{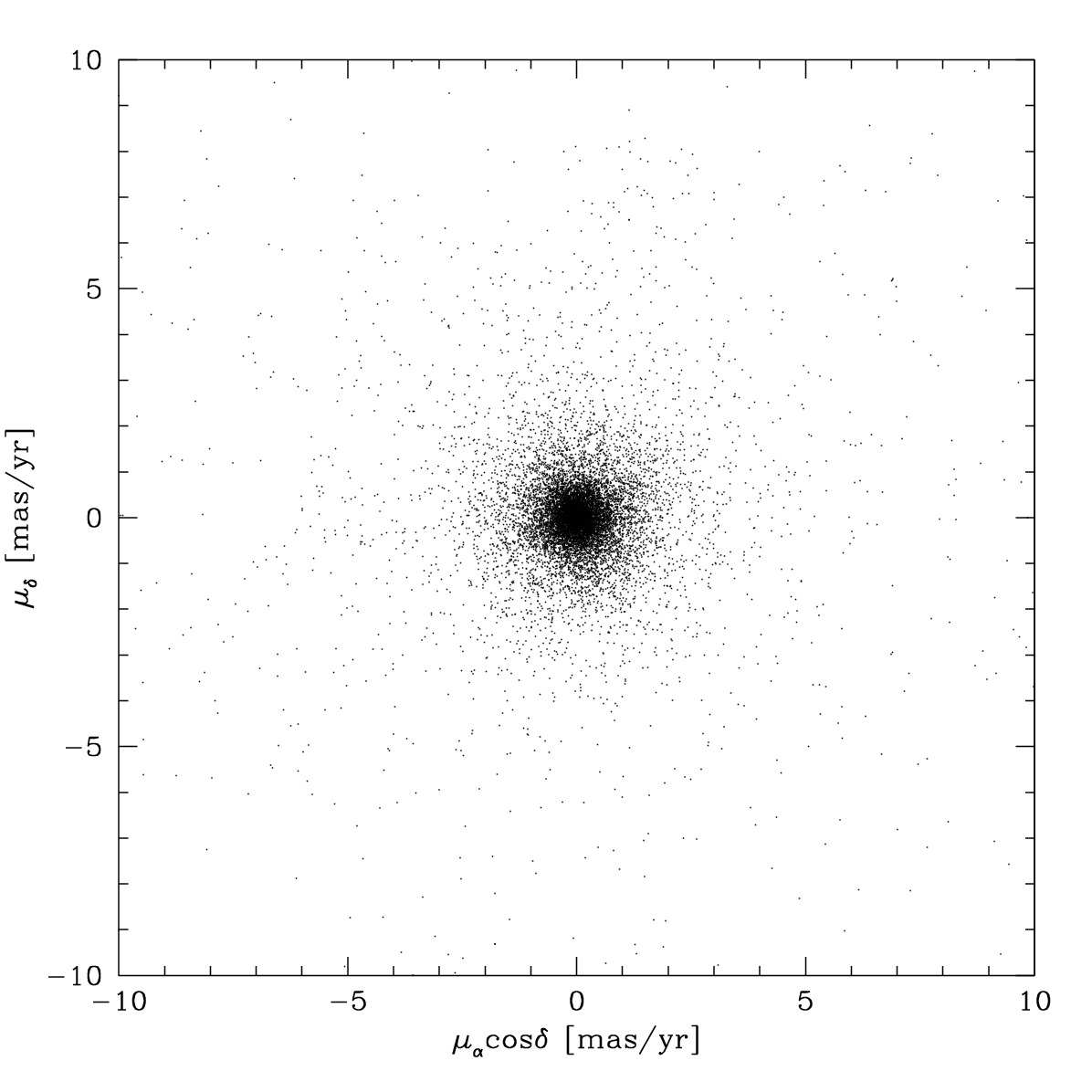}
    \caption{VPD for M30.}
    \label{fig:m30vpd}
\end{figure}

\begin{figure}
	\includegraphics[width=\columnwidth]{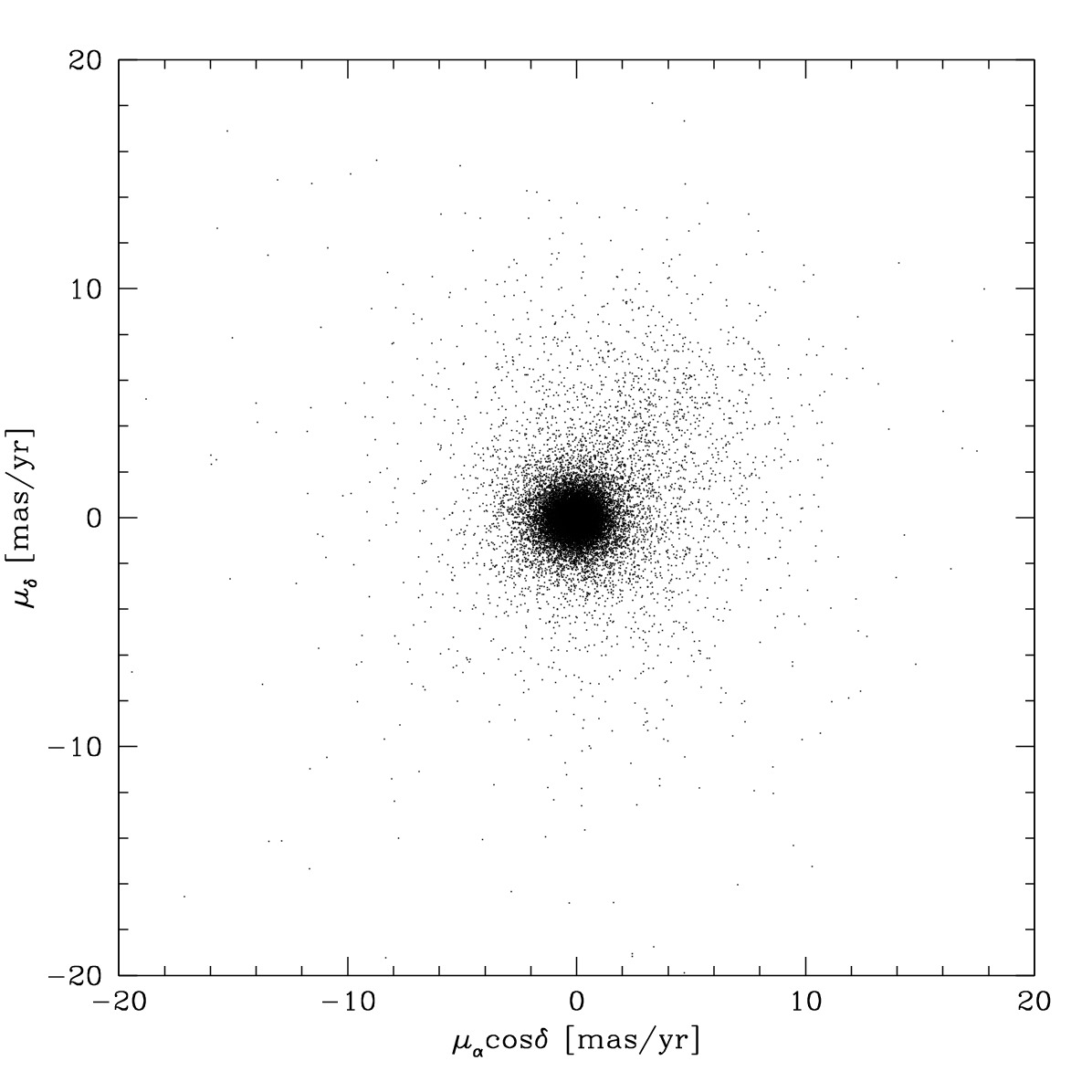}
    \caption{VPD for M10.}
    \label{fig:m10vpd}
\end{figure}

\begin{figure}
	\includegraphics[width=\columnwidth]{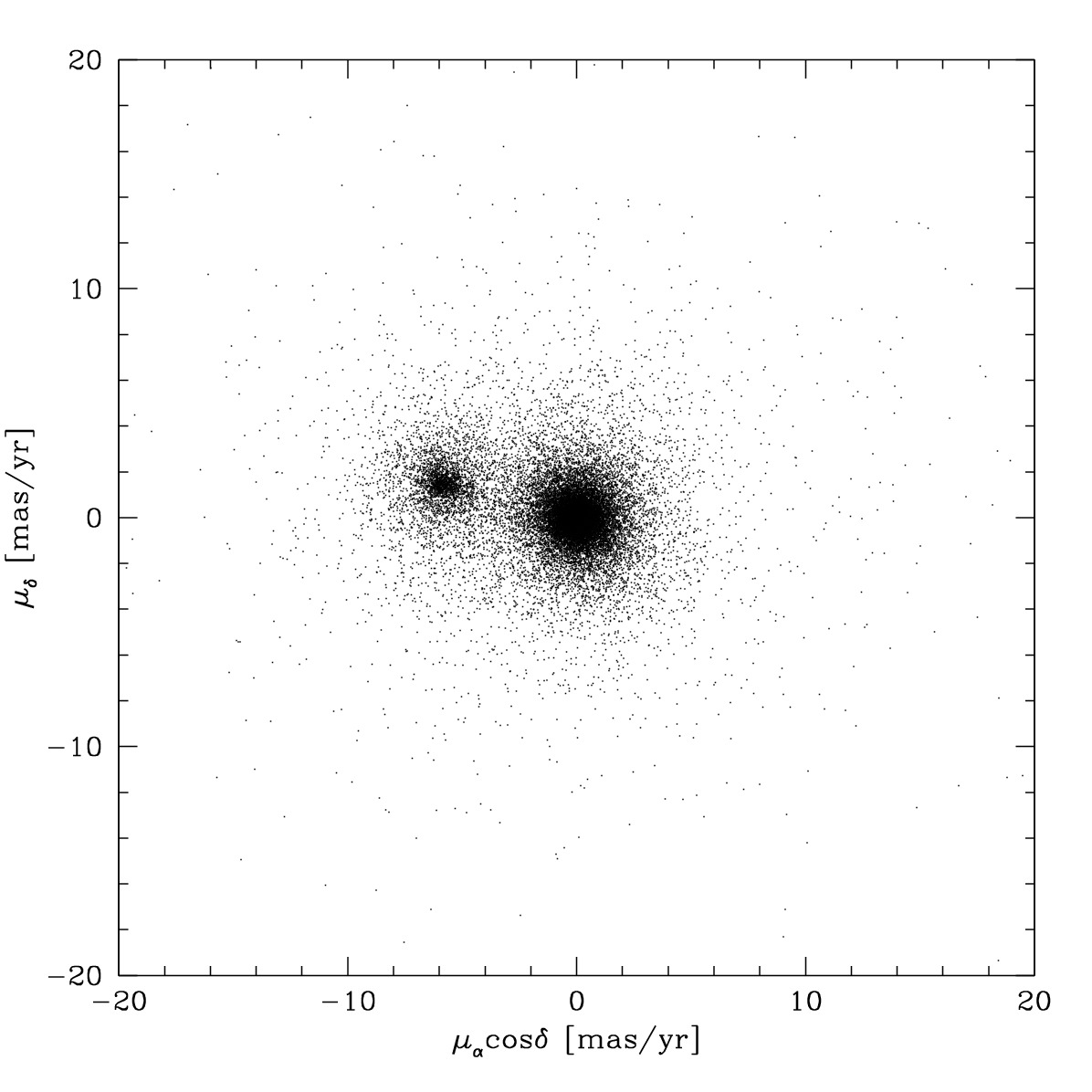}
    \caption{VPD for NGC~362.}
    \label{fig:ngc362vpd}
\end{figure}

\begin{figure}
	\includegraphics[width=\columnwidth]{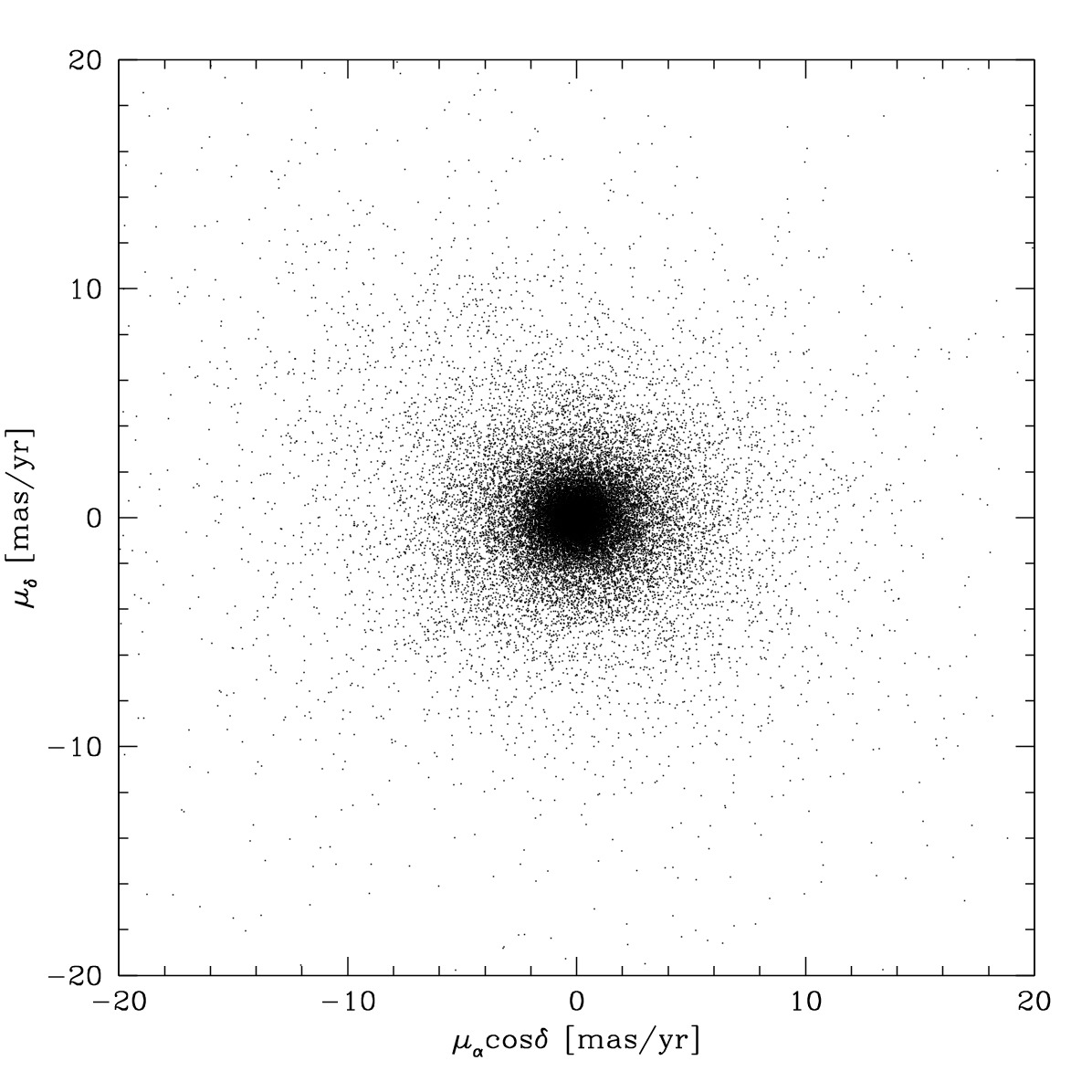}
    \caption{VPD for M5.}
    \label{fig:m5vpd}
\end{figure}

\begin{figure}
	\includegraphics[width=\columnwidth]{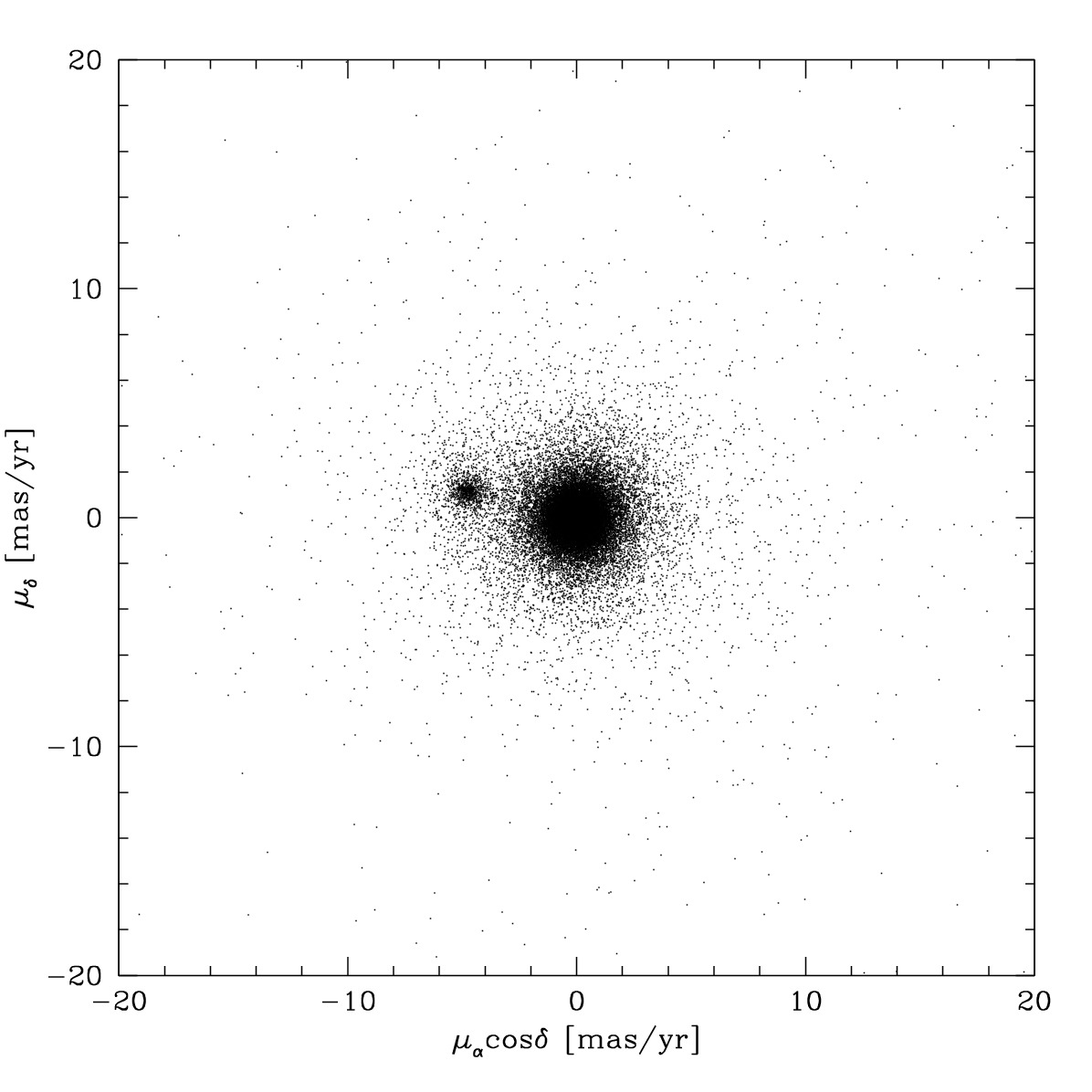}
    \caption{VPD for 47~Tuc~E.}
    \label{fig:47tucEvpd}
\end{figure}

\begin{figure}
	\includegraphics[width=\columnwidth]{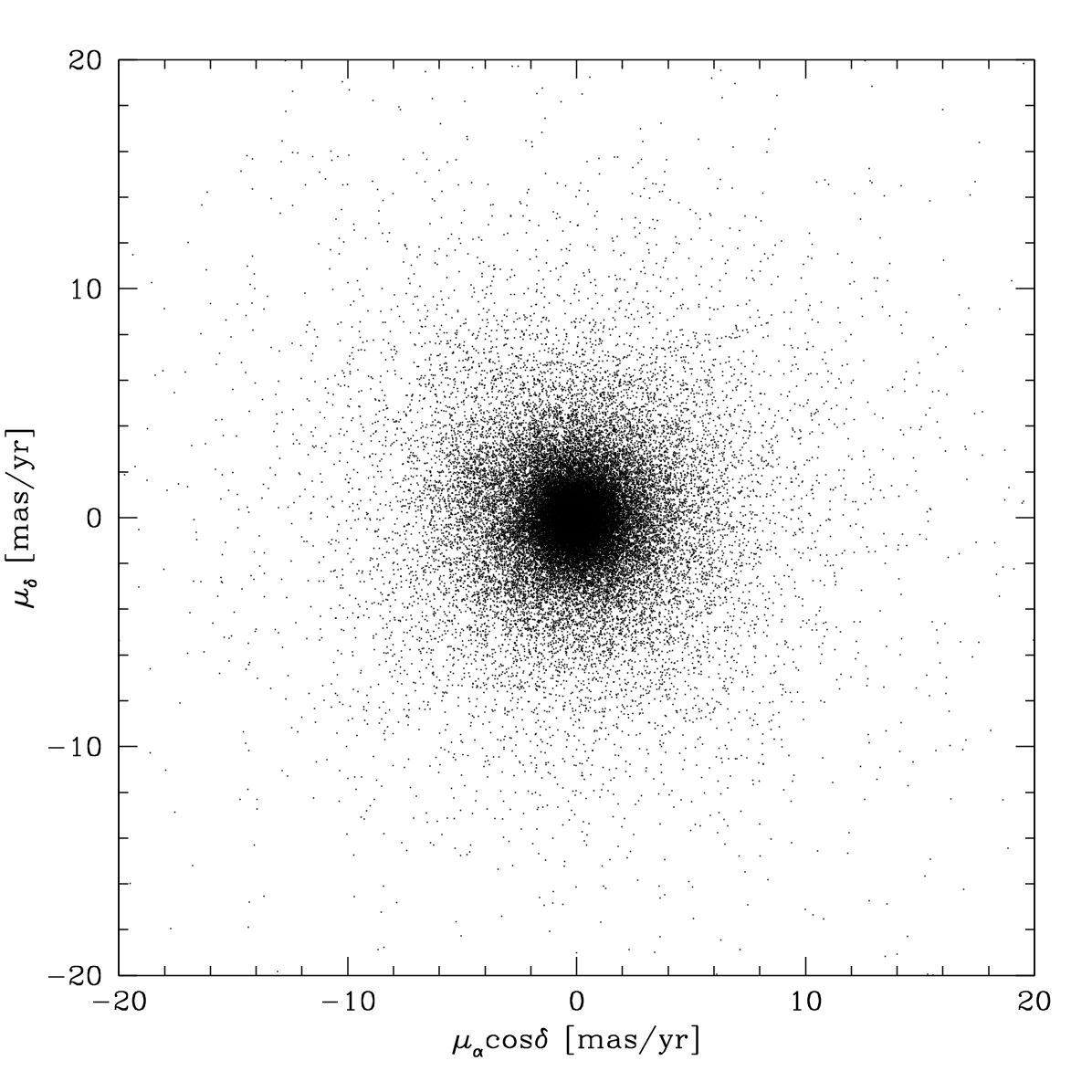}
    \caption{VPD for 47~Tuc~W.}
    \label{fig:47tucWvpd}
\end{figure}

Fig.~\ref{fig:m12vpd}--\ref{fig:47tucWvpd} present VPDs for all GCs in this study.

\bsp	
\label{lastpage}
\end{document}